\documentclass{aa}  

\usepackage{graphicx}
\usepackage{txfonts}
\usepackage{color}
\definecolor{mygreen}{RGB}{84,141,40}
\usepackage{siunitx}
\begin{document}

   \title{Constraining nearby substellar companion architectures using High Contrast Imaging, Radial Velocity and Astrometry}
   \titlerunning{Constraining nearby substellar companion architectures using HCI, RV, and Astrometry}

   \author{L. F. Sartori
          \inst{1}\fnmsep\thanks{E-mail: lia.sartori@phys.ethz.ch}
          \and
          M. J. Bonse \inst{1}
          \and
          Y. Li \inst{2}
          \and
          F. A. Dannert \inst{1}
          \and
          S. P. Quanz \inst{1,3}
          \and
          C. Lovis \inst{4}
          \and 
          A. Boehle \inst{1}
          }

   \institute{ETH Zurich, Institute for Particle Physics and Astrophysics, Wolfgang-Pauli-Strasse 27, CH-8093 Zurich, Switzerland
         \and
             Department of Astronomy, University of Michigan, Ann Arbor, MI 48109, USA
        \and 
            ETH Zurich, Department of Earth and Planetary Sciences, Sonneggstrasse 5, 8092 Zurich, Switzerland
        \and
            D\'{e}partement d'Astronomie, Universit\'{e} de Gen\`{e}ve, CH-1290 Versoix, Switzerland
             }

   \date{Received xxx; accepted xxx}

  \abstract
   {Nearby stars offer prime opportunities for exoplanet discovery and characterization through various detection methods. By combining high-contrast imaging (HCI), radial velocity (RV), and astrometry, it is possible to better constrain the presence of substellar companions, as each method probes different regions of their parameter space. A detailed census of planets around nearby stars is essential to guide the selection of targets for future space missions seeking to identify Earth-like planets and potentially habitable worlds. In addition, the detection and characterisation of giant planets and brown dwarfs is crucial for understanding the formation and evolution of planetary systems.}
   {We aim to constrain the possible presence of substellar companions for 7 nearby M-dwarf stars (GJ~3325, GJ~1125, GJ~367, GJ~382, GJ~402, GJ~465 and GJ~357) using a combination of new SPHERE/\textit{H2} HCI and archival RV and astrometric data. We investigate how combining these techniques improves the detection constraints for giant planets and brown dwarfs compared to using each method individually.
}
   {For each star and each data set, we compute the mass limits as a function of semi-major axis or projected separation using standard techniques. We then use a Monte Carlo approach to assess the completeness of the companion mass / semi-major axis parameter space probed by the combination of the three methods, as well as by the three methods independently.
}
   {Our combined approach significantly increases the fraction of detectable companions. We quantify improvements of up to $\sim 60\%$ over RV alone (improvement achieved for GJ~3325), $\sim 50\%$ over HCI alone (for GJ~3325), and $\sim 12\%$ over astrometry alone (for GJ~367). Although no new companion was detected, we could place stronger constraints on potential substellar companions.
}
   {The combination of HCI, RV and astrometry provides significant improvements in the detection of substellar companions over a wider parameter space. Applying this approach to larger samples and lower-mass companions will help constraining the search space for future space missions aimed at finding potentially habitable or even inhabited planets.}


   \maketitle
\section{Introduction}

In the coming years and decades, large scale exoplanet imaging surveys and space missions will search for exoplanets and brown dwarfs around hundreds of stars, with a focus on the Solar neighbourhood. These surveys will be performed with instruments such as the ERIS/NIX imager at the VLT (\citealt{Dubber2022}; \citealt{Davies2023}), the Mid-infrared ELT Imager and Spectrograph (METIS, \citealt{Quanz2015}; \citealt{Brandl2021}) or the Planetary Camera and Spectrograph (PCS, \citealt{Kasper2021}) at the ELT, as well as by space missions such as the Large Interferometer for Exoplanets (LIFE, \citealt{Quanz2018,Quanz2022}) and the Habitable Worlds Observatory (HWO, \citealt{LUVOIR2019}; \citealt{Gaudi2020}; \citealt{National2021}). The goal of such efforts will be to determine the architecture and demographics of planetary systems, investigate planet formation and evolution, and ultimately look for life beyond the Solar System. In this context, constraining the architecture of substellar companions, particularly giant planets and brown dwarfs, is essential for advancing our understanding of planetary system formation and evolution (e.g., \citealt{Vorobyov2013}; \citealt{Kley2017}; \citealt{Matsukoba2023}). These objects, which bridge the gap between the most massive planets and the least massive stars, can form through distinct mechanisms such as core accretion (e.g., \citealt{Goldreich1973}; \citealt{Pollack1996}; \citealt{Ida2004}; \citealt{Mordasini2009,Mordasini2012}) or gravitational instability (e.g., \citealt{Boss1998}; \citealt{Vorobyov2013}; \citealt{Mayer2002}; \citealt{Boss2006}), each resulting in characteristic mass and orbital distributions (e.g., \citealt{Janson2011}; \citealt{Rameau2013}; \citealt{Nielsen2019}). By characterising their statistics, both in terms of population number and orbital arrangement, we can gain crucial insights into the relative importance of these formation pathways. Additionally, the presence and configuration of such massive companions can significantly influence the dynamical evolution of planetary systems, affecting the stability and habitability of inner, terrestrial planets (e.g., \citealt{Konopacky2016}; \citealt{Wang2018b}; \citealt{Nielsen2020}; \citealt{Bitsch2015}; \citealt{Izidoro2015}). Consequently, accurate constraints on the occurrence and orbital properties of giant planets and brown dwarfs in nearby systems are crucial not only for population studies but also for guiding future observational campaigns aimed at detecting potentially habitable exoplanets around these stars.

Given the limited availability of observing time, the large-scale surveys and missions mentioned above will hardly be able to conduct blind searches for substellar companions. Therefore, stellar target samples must be carefully pre-selected based on current knowledge of planetary system statistics and architectures, as well as theoretical models of their formation and evolution (e.g., \citealt{Dubber2022}; \citealt{Tuchow2024}; \citealt{Menti2024}). This pre-selection is primarily informed by the characteristics of known exoplanetary systems and accepted formation models (e.g., \citealt{Meyer2018}; \citealt{Nielsen2019}; \citealt{Vigan2021}; \citealt{Barbieri2023}). Several recent imaging programs have also demonstrated the effectiveness of using RV trends and/or high SNR astrometry to guide target selection, thereby increasing the likelihood of detecting massive companions (e.g., \citealt{Calissendorff2018}; \citealt{Nowak2020}; \citealt{Bonavita2022}; \citealt{Mesa2022,Mesa2023}; \citealt{Hinkley2023}; \citealt{Barbato2025}).
However, despite the thousands of companions discovered to date, the existing sample is highly biased due to the intrinsic limitations of the various detection methods (e.g., \citealt{Perryman2018}). Radial velocity (RV) observations are most effective at detecting massive companions on close orbits, such as hot Jupiters, or smaller planets in even tighter configurations. High-contrast imaging (HCI) is sensitive to large, young, and hot planets on wide orbits. Astrometry, on the other hand, is particularly suited for detecting massive companions at intermediate orbital distances, especially around nearby stars. Each method therefore explores a distinct region of the companion mass and semi-major axis parameter space, which must be accounted for when characterising the demographics of planetary systems. By combining results from multiple detection techniques, a broader and more representative picture of the population of substellar companions can be obtained.

In addition to this, existing archival data, which often remain unpublished due to non-detections, represent an underutilized resource. Properly analyzing these data to establish upper mass limits and detection thresholds can provide stringent constraints on the parameter space for potential substellar companions (e.g., \citealt{Bonfils2013}; \citealt{Boehle2019}; \citealt{Hurt2021}; \citealt{Bonse2024}; \citealt{Dietrich2024}). These non-detections are valuable for delineating regions of parameter space that could or cannot host (yet) undetected companions, thereby refining our understanding of the true architecture of nearby planetary systems.

The combination of complementary detection techniques to achieve a more comprehensive understanding of planetary and stellar systems has been explored in recent studies (e.g., \citealt{Boehle2019}; \citealt{Dulz2020}; \citealt{Lagrange2020}; \citealt{Wood2021}; \citealt{Dietrich2024}; \citealt{Gratton2024,Gratton2025}; \citealt{Barbato2025}). For instance, \citet{Boehle2019} combined HCI observations from VLT/NACO (\citealt{Rousset2003}; \citealt{Lenzen2003}) with archival RV data for six nearby stars, showing that HCI can increase the completeness for certain masses and semi-major axes up to $99\%$ compared to RV alone. Expanding on this approach, \citet{Wood2021} incorporated Gaia astrometric data, utilizing the renormalised unit weight error (RUWE; \citealt{Lindegren2018}) to constrain the properties of potential stellar companions in binary systems. These studies highlight the utility of multi-technique analyses in refining our constraints on the architectures of nearby planetary systems, by leveraging the strengths of each detection method and mitigating their individual limitations.

In this study, we combine newly obtained HCI observations of 7 nearby stars acquired with the Spectro-Polarimetric High-contrast Imager for Exoplanet REsearch (SPHERE; \citealt{Bezuit2019}; \citealt{Vigan2010a}) with archival RV data and proper motion anomalies (PMa) derived from the Hipparcos (\citealt{Perryman1997}; \citealt{vanLeeuwen2007}) and Gaia (E)DR3 (\citealt{GaiaCollaboration2016, GaiaCollaboration2020, GaiaCollaboration2021, GaiaCollaboration2021a}) catalogues (\citealt{Kervella2021, Kervella2022}). While the renormalised unit weight error (RUWE) parameter is available for all Gaia sources and provides a broad indicator of astrometric anomalies, the PMa offers more direct and quantitative constraints on the properties of potential companions, making it particularly suitable for this study. This work is conducted as part of the Planetary Systems In Our Neighbourhood (PSION) project, which aims to identify and characterise companions around some of the closest stars to the Sun, in order to constrain the architectures of their planetary systems and to prioritise targets for future observational campaigns (e.g., \citealt{Boehle2019}; \citealt{Sartori2023}; \citealt{Menti2024}). Although the present results are limited by the small sample size, this study demonstrates the effectiveness of combining HCI, RV, and astrometric data for constraining the parameter space of (substellar) companions. It can also can be regarded as a proof-of-concept for future more comprehensive studies as the methodology can be easily applied to larger samples, providing valuable input for defining target lists for upcoming space missions such as LIFE and the HWO.

The paper is organised as follows. In Section \ref{sec:sample} we describe the sample selection, the available data (new and archival) and the performed data reduction. In Section \ref{sec:analysis} we illustrate the analysis performed to compute the constraints for each technique alone, and how we combine them to investigate the architecture of the systems. In Section \ref{sec:results} we summarise the results and compare them with the expectation from population models. Finally, we discuss our findings in Section~\ref{sec:dis}.

\section{Sample and data}\label{sec:sample}

\subsection{Target stars}\label{sec:sample_sel}
Our sample consists of seven nearby ($d < 10$pc) M dwarf stars for which we obtained new high-contrast imaging in the \textit{H2} and \textit{H3} bands with the SPHERE instrument (Program ID 0104.C$-$0336, PI A. Boehle, and Program ID 112.25G2, PI L. Sartori. See Section~\ref{sec:HCI_data} for details): GJ~3325, GJ~1125, GJ~367, GJ~382, GJ~402, GJ~465 and GJ~357. Details about these stars are summarised in Table \ref{tab:star_par}. Since our primary goal is to demonstrate the potential of the combined method rather than to perform a detailed stellar characterization, we adopt literature values for stellar parameters where available, acknowledging the resulting inhomogeneities.

As part of the PSION project, the targets were chosen from a volume-limited sample of all known single, dwarf stars within 10 pc with spectral types from F$0$ to M$4$ (94 stars in total). To compile the total sample, we first compared the predicted detection limits in the companion mass/semi-major axis parameter space from coronagraphic IRDIS H2/H3 dual band imaging\footnote{Based on the ESO SPHERE/IRDIS ETC calculator \url{https://www.eso.org/observing/etc/}} to those derived from archival long-term radial velocity data, mostly from the High Accuracy Radial velocity Planet Searcher (HARPS, \citealt{Pepe2002}; \citealt{Mayor2003}) and High Resolution Echelle Spectrometer (HIRES, \citealt{Vogt1994}) instruments (see Section~\ref{sec:RV_data}), and selected the stars for which the SPHERE observations were predicted to be sensitive enough to substellar companions that are currently missed by the radial velocity data. Specifically, we predicted the HCI contrast curves using the ESO Exposure Time Calculator, while for the RV limits we followed the same procedure as described in Section \ref{sec:RV_ml}. We then checked that their apparent magnitude is bright enough to enable the use of the wavefront sensor of the SPHERE extreme AO system ($r < 11$~mag, \citealt{Fusco2006}). The final sample consisted of the aforementioned 7 M dwarf stars with no previous archival SPHERE observations. We obtained the requested observations for 5 of them during the ESO period P104, while 3 were observed during the ESO period P112.

Among the stars in our sample, only GJ~367 and GJ~357 have previously detected exoplanet companions. Recent work based on TESS and HARPS data for GJ~367 (\citealt{Lam2021}; \citealt{Goffo2023}) confirmed the presence of three planets with masses m$_{\rm b} = 0.6 \pm 0.083$ M$_\oplus$, m$_{\rm c} = 4.08 \pm 0.30$ M$_\oplus$ and m$_{\rm d} = 5.93 \pm 0.45$ M$_\oplus$, and orbiting periods P$_{\rm b} = 0.3219225 \pm 0.0000002$ days, P$_{\rm c} = 11.543 \pm 0.005$ days and P$_{\rm d} = 34.39 \pm 0.06$ days. These planets are detected also in our archival RV timeseries, and we discuss in Section \ref{sec:RV_data} how we treated the corresponding signal in our analysis. For GJ~357, the analysis of TESS and CARMENES data (\citealt{Luque2019}) confirmed the presence of three planets with masses m$_{\rm b} = 1.84 \pm 0.031$ M$_\oplus$, m$_{\rm c} = 3.40 \pm 0.46$ M$_\oplus$ and m$_{\rm d} = 6.1 \pm 1.0$ M$_\oplus$, and orbiting periods P$_{\rm b} = 3.93$ days, P$_{\rm c} = 9.12$ days and P$_{\rm d} = 55.7$ days. We have treated these data in the same way as described for GJ~367. Because of the short periods and small semi-major axes, none of these planets is visible in our HCI data.

\begin{table*}
    \centering
    \begin{tabular}{lccccccl}
    \hline
        Target & Spectral Type & Distance & Stellar Mass & Stellar Age & T$_{\rm{eff}}$ & V & References\\
        ~ & ~ & [pc] & [M$_\odot$] & [Gyr] & [K] & [mag] & ~ \\
    \hline
        GJ 3325 & M3V & 9.23 & 0.290 $\pm$ 0.014 & 5.6 & 3401 & 11.73 & [1],[2],[3],[4],[5] \\
        GJ 1125 & M3.5V & 9.90 & 0.303 & 4.98 & 3089 & 11.71 & [4],[6],[7],[8],[9] \\
        GJ 367 & M1.0 & 9.41 & 0.455 & 8.65 & 3283 & 9.98 & [4],[6],[8],[10]  \\
        GJ 382 & M2V & 7.70 & 0.508 & 5.6 & 3359 & 9.26 & [3], [6],[8],[11],[12] \\
        GJ 402 & M4V & 6.97 & 0.282 $\pm$ 0.014 & 7.5 & 3001 & 11.67 & [2],[6],[11],[13],[14] \\
        GJ 465 & dM2.0 & 8.88 & 0.274 $\pm$ 0.014 & 1.82 & 3219 & 11.27 & [2],[4],[6],[9],[15] \\
        GJ 357 & M2.5V & 9.44 & 0.346 $\pm$ 0.20 & & 3204 & 10.91 & [6],[4],[8],[9],[16] \\
    \hline
    \end{tabular}
    \caption{Stellar properties of the seven targets of this paper. \\ {\bf{References}}: \textit{Spectral Type: } [1] \cite{Henry2002}; [4] \cite{Koen2010}; [7] \cite{Lepine2013}; [11] \cite{Kirkpatrick1991}; [15] \cite{Schweitzer2019}; [16] \cite{Gray2006}; \\ \textit{Stellar Mass: } [2] \cite{Winters2021}; [8] \cite{Stassun2019}; \\
    \textit{Stellar Age: } [3] \cite{Yee2017}; [9] \cite{Queiroz2023}; [10] \cite{Lu2024}; [13] \cite{Brandt2015}; \\ 
    \textit{Effective Temperature: } [5] \cite{Maldonado2020}; [6] \cite{GaiaCollaboration2022}; \\
    \textit{V mag: } [4] \cite{Koen2010}; [12] \cite{Kirga2012}; [14] \cite{Landolt1992}}
    \label{tab:star_par}
\end{table*}

\subsection{High-contrast imaging}\label{sec:HCI_data}

We obtained SPHERE high contrast imaging data of the seven target stars between January and March 2020, and between January and March 2024. Observations are taken in the IRDIFS mode, which employs simultaneously the infrared dual-band imager and spectrograph (IRDIS, \citealt{Dohlen2008}) in its dual-band imaging mode (DBI, \citealt{Vigan2010a}) and the integral field spectrocopic mode (IFS, \citealt{Claudi2008}; \citealt{Mesa2015}). For DBI we used the {\textit{H2H3}} filter pair ($\lambda_{H2} = 1.593~\mu$m, $\Delta\lambda_{H2} = 0.052~\mu$m, $\lambda_{H3} = 1.667~\mu$m, $\Delta\lambda_{H3} = 0.056~\mu$m) in pupil-stabilized mode to allow angular differential imaging (ADI, \citealt{Marois2006}). This filter pair is ideal for searching for planets and eventually studying their atmospheres, as it covers one of the main methane absorption features (e.g., \citealt{Bonnefoy2018}). For simultaneous IFS observations, we used the {\textit{YJ}} mode ($\lambda = 0.95 - 1.35 \mu$m, spectral resolution $R \sim 54$), but these data are not discussed in this paper, as they do not add additional information useful for our analysis (no planet is detected). The observations were taken with the ALC$\_$YJH$\_$S Lyot coronagraph (\citealt{Soummer2005}) which is optimised for the $H$-band and has an inner working angle IWA $\sim 0".15$\footnote{From the SPHERE User Manual, \url{https://www.eso.org/sci/facilities/paranal/instruments/sphere/doc.html}}. To obtain a model of the point spread function (PSF), the stars were observed before and after the science sequence without the coronagraph but with an additional neutral density filter, ND2.0, with a transmission of $\sim 1 \%$ ($1.2 \%$ for {\textit{H2}} and $1.7 \%$ for {\textit{H3}}) of the total bandwidth\footnote{For GJ382, the neutral density filter ND3.5 was used instead of the ND2.0 one.}. Additional information about the observations are provided in Table \ref{tab:hci_obs}.

We obtained the raw files from
the ESO archive and performed a standard reduction using the \texttt{vlt-sphere} package (\citealt{Vigan2020}), which relies on the official ESO pipeline (\citealt{CPL2014}). The standard calibration  includes dark and background subtraction, flat field and bad pixels correction, wavelength calibration (for the IFS data) and star centering. We then used the state-of-the-art pipeline for direct-imaging data reduction and analysis \texttt{PynPoint} \citep{Stolker2019} to divide the reduced DBI data cubes into single bands, median combine the point spread function (PSF) cubes obtained before and after the science observations, as well as perform Angular Differential Imaging (ADI) and Principal Component Analysis (PCA) (\citealt{AmaraQuanz2012}, \citealt{Soummer2012}) PSF subtraction. The residual images for H2 are shown in Fig. \ref{fig:H2_residuals}. No companion signal is detected in these images. We could therefore compute the contrast curves without subtracting any signal, as described in Section \ref{sec:cc}.

\begin{table*}
\centering
\begin{minipage}{0.65\textwidth}
\begin{tabular}{llcc}
\hline
    Target & UT date & FoV rot. & Seeing$^{\rm{(a)}}$\\
    ~ & ~ & [deg] & [arcsec]\\
\hline
    GJ 3325 & 2020-01-19 & 78.5 & $ 0.50 \pm 0.04 $ \\
    GJ 1125 & 2020-02-07 & 31.9 & $ 0.71 \pm 0.22 $ \\
    GJ 367  & 2020-02-08 & 30.4 & $ 0.64 \pm 0.13 $ \\
    GJ 382  & 2020-02-22 & 37.3 & $ 0.76 \pm 0.10 $ \\
    GJ 402  & 2024-01-21 & 25.5 & $ 0.33 \pm 0.05 $ \\
    GJ 465  & 2024-03-07 & 94.2 & $ 0.54 \pm 0.06 $ \\
    GJ 357  & 2024-01-20 & 126.5 & $ 0.47 \pm 0.08 $ \\
\hline
\end{tabular}
\end{minipage}%
\hfill
\begin{minipage}{0.32\textwidth}
\vspace{-1em} %
\begin{tabular}{ll}
\hline
IRDIS filter & \textit{H2H3} \\
IFS band & \textit{YJ} \\
IRDIS DIT$^{\rm{(b)}}$ $\times$ NDIT$^{\rm{(c)}}$ & $8 \times 2$ s \\
IRDIS T$_{\rm{exp}}$ & 46.9 min \\
IFS DIT $\times$ NDIT & $32 \times 1$ s \\
IFS T$_{\rm{exp}}$ & 51.7 min \\
\hline
\end{tabular}
\end{minipage}

\vspace{0.5em}
\caption{Summary of the new SPHERE observations of the seven targets. $^{\rm{(a)}}$ Mean seeing during observations. $^{\rm{{(b)}}}$ Detector Integration Time in seconds. $^{\rm{{(c)}}}$ Number of frames per dithering position.}
\label{tab:hci_obs}
\end{table*}

\subsection{Radial velocity}\label{sec:RV_data}

The archival HARPS and HIRES radial velocity measurements used in this work are summarised in Table~\ref{tab:astr}. The data were processed and analysed within the Data \& Analysis Center for Exoplanets (DACE) platform, a facility hosted by the University of Geneva which is dedicated to extrasolar planets data visualisation, exchange and analysis\footnote{\url{https://dace.unige.ch/dashboard/index.html}}. The RV and stellar activity indicators time series available in DACE are computed by cross-correlating the calibrated spectra with stellar templates of the corresponding stellar class using state-of-the-art pipelines\footnote{Specifically, the RVs and activity indicators in DACE are derived using the Data Reduction Software (DRS) for HARPS, which employs the cross-correlation function (CCF) method (\citealt{Pepe2002}; \citealt{Mayor2003}), and the Keck Observatory HIRES pipeline for HIRES, which also uses the CCF method (\citealt{Vogt1994}; \citealt{Isaacson2010}).}. We retrieved the available RV time series using the DACE python APIs, and analysed them in a similar way as described in \cite{Sartori2023}.
In summary, for each target we subtracted a linear offset arising from the recession velocity of the star itself and from instruments cross-calibration, added the instrumental jitter to the error bars, and then computed the generalised Lomb-Scargle periodograms (\citealt{Cumming2008}, with formalism from \citealt{Zechmeister2009}). For all stars except GJ~367 and GJ~357, the obtained periodograms do not show any peak indicating the presence of an orbiting planet at any of the probed periods. We therefore used the obtained RV time series without subtracting any signal to compute the mass limits, as described in Section~\ref{sec:RV_ml}. For GJ~367 and and GJ~357 instead we were able to reproduce the analysis in the discovery papers our Keplerian fit returned values consistent with those in \cite{Goffo2023} \citealt{Luque2019}, respectively. We subtracted the signal arising from the known planets before computing the mass limits.

We note that in order to ensure a consistent analysis across our sample and to avoid the need for re-reducing or homogenizing heterogeneous datasets, we limit our RV analysis to time series available through the DACE platform. While this excludes some additional measurements available in the literature (e.g., CARMENES, PFS, or UVES for GJ 357), we consider the available data sufficient for the illustrative purpose of this work.

\begin{table*}
    \centering
    \begin{tabular}{lccccccc}
    \hline
        Target & \multicolumn{4}{c}{No. of measurements} & $\Delta t_{\rm{RV}}$$^{\rm{(a)}}$ & $a$ for $\Delta t_{\rm{RV}}$$^{\rm{(b)}}$ & $\sigma_{\rm{RV}}$ \\
        \cline{2-5}
        ~ & HIRES & HARPS 03 & HARPS 15 & Total & [days] & [AU] & [ms$^{-1}$] \\
    \hline
        GJ 1125 & 19 & 0 & 0 & 19 & 4090 & 3.4 & 17.3\\ %
        GJ 367 & 0 & 25 & 346 & 371 & 6391 & 5.2 & 3.12\\ %
        GJ 382 & 96 & 34 & 0 & 130 & 6068 & 5.2 & 4.3\\ %
        GJ 3325 & 21 & 9 & 1 & 31 & 6961 & 4.7 & 6.2\\ %
        GJ 402 & 24 & 6 & 5 & 45 & 5864 & 4.2 & 7.6\\
        GJ 465 & 0 & 23 & 20 & 43 & 5140 & 3.8 & 6.2\\
        GJ 357 & 61 & 53 & 0 & 114 & 5884 & 4.5 & 4.4\\
    \hline
    \end{tabular}
    \caption{Summary of the archival radial velocity (RV) observations available for four of the five targets in our sample. $^{\rm{(a)}}$ Total time baseline for the RV measurements. $^{\rm{(b)}}$ Semi-major axis $a$ correspondent to a revolution with period equal to the total time baseline $\Delta t_{\rm{RV}}$, given the stellar masses in Table \ref{tab:star_par}.}
    \label{tab:RV_par}
\end{table*}

\subsection{Astrometry}

Seven of our targets are present in both the Hipparcos (\citealt{Perryman1997}; \citealt{vanLeeuwen2007}) and Gaia EDR3 (\citealt{GaiaCollaboration2016,GaiaCollaboration2020,GaiaCollaboration2021,GaiaCollaboration2021a}) catalogues. For this study we directly retrieved the proper motion anomaly (PMa) values from \citealt{Kervella2021}. Details about the astrometric data are given in Table~\ref{tab:astr}. We have verified that the observations are not saturated (G mag $>> 3$) and that all Gaia (E)DR3 measurements have low RUWE ($<1.2$ for the entire sample) and low excess noise ($\epsilon_i < 0.35$) values, meaning that the measurements are reliable for our analysis and do not indicate the presence of stellar companions orbiting at large distances. %

\begin{table*}
    \centering
    \begin{tabular}{llcccc}
    \hline
        Target & GAIA DR3 ID & HIPPARCOS ID & $\Delta v_{{\rm{T, GDR3}}}$ $^{\rm{{(a)}}}$ & $r_{\rm{orb}}$ for $\delta t_{\rm{GDR3}}$ $^{\rm{{(b)}}}$ \\
        ~ & ~ & ~ & [ms$^{-1}$] & [AU] \\
    \hline
        GJ 1125 & 3840711729206820000 & 46655 & $18.7 \pm 7.3$ & 1.36 \\
        GJ 367  & 5412250540681250560 & 47780 & $21.13 \pm 5.55$ & 1.56 \\
        GJ 382  & 3828238392559860992 & 49986 & $3.78 \pm 2.33$ & 1.62 \\
        GJ 3325 & 2976850598890749824 & 23512 & $1.2 \pm 3.5$ & 1.31 \\
        GJ 402 & 3864615459886222464 & 53020 & $2.9 \pm 4.1$ & 1.309 \\
        GJ 465 & 3519785523672576384 & 60559 & $2.1 \pm 4.0$ & 1.29 \\
        GJ 357 & 5664814198431308288 & 47103 & $1.6 \pm 3.2$ & 1.42 \\
    \hline
    \end{tabular}
    \caption{Summary of the astrometric measurements from \protect\cite{Kervella2021} for the four targets present in both the Hipparcos and Gaia EDR3 catalogues. $^{\rm{{(a)}}}$ Norm of the tangential proper motion anomaly (PMa) vector converted to linear velocity using the Gaia EDR3 parallax. $^{\rm{{(b)}}}$ Radius of an orbit with period corresponding to the Gaia DR3 observing window $\delta t_{\rm{GDR3}} = 1038$d given the stellar mass of the target.}
    \label{tab:astr}
\end{table*}

\section{Analysis}\label{sec:analysis}
\subsection{Contrast curves and mass limits from HCI}\label{sec:cc}

\begin{figure*}
\includegraphics[width=\textwidth]{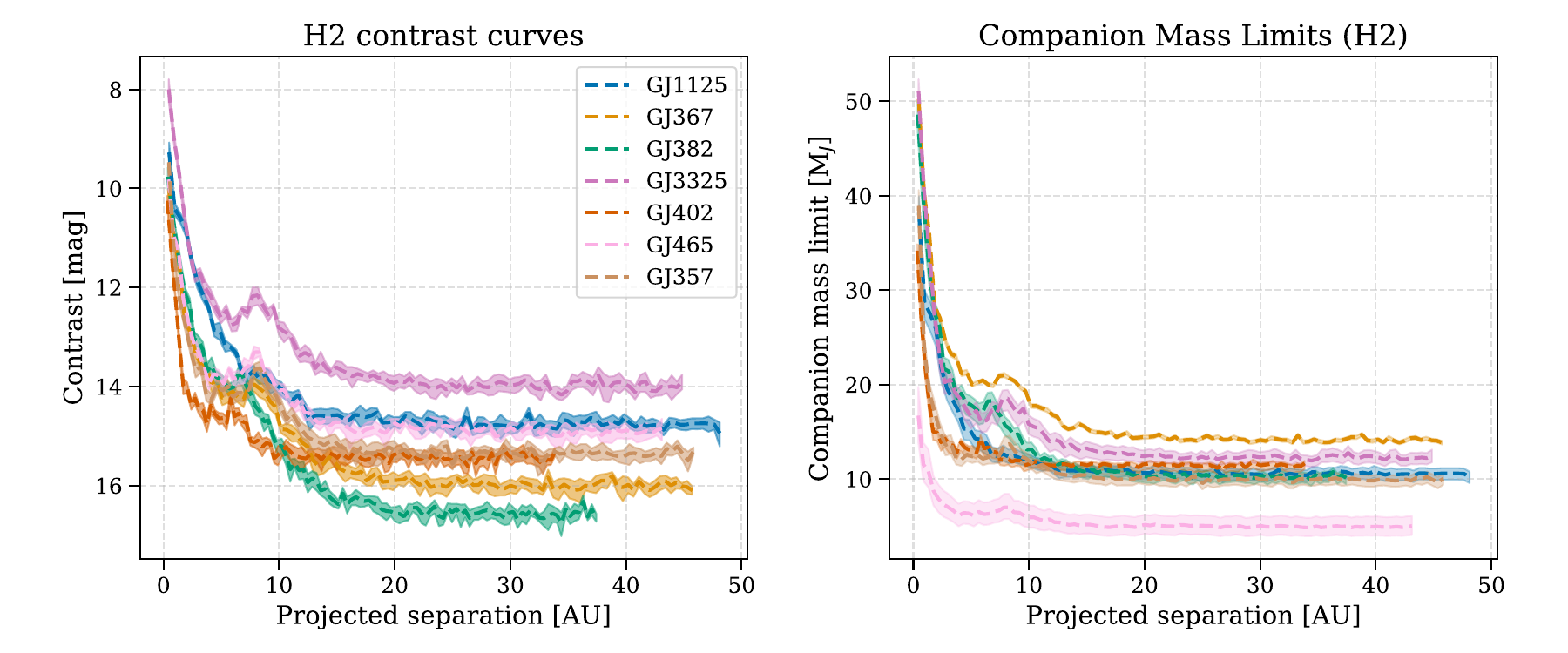}
\caption{\textit{Left:} Contrast curves computed from the SPHERE H2 high contrast images. The rise in most contrast curves at around 1 arcsec (converted to physical scale) corresponds to the limit of the AO correction (0.8 arcsec, \citealt{Fusco2006,Fusco2016}). The reached contrast in the H2 band and H3 band (not plotted here) is almost identical. \textit{Right:} Mass limits derived from the SPHERE H2 contrast curves in above using the AMES-Cond evolutionary models (\citealt{Baraffe2003}). The shaded regions show the possible spread due to the uncertainty on the stellar ages. We note that the mass limits obtained in the H2 filter are a factor $\sim2$ better than in the H3 filter.}
\label{fig:HCI_cc_ml}
\end{figure*}

We computed the analytical contrast curves for each star and for both the {\textit{H2}} and {\textit{H3}} filters separately using the \texttt{python} package \texttt{applefy}\footnote{\url{https://applefy.readthedocs.io/}} \cite{Bonse2023}. 
To account for the attenuation of the signal due to post-processing with PCA, we inserted artificial planets into the data using the unsaturated off-axis stellar PSF.
The artificial planets were inserted at different separations every $1 \lambda /D$ and at 6 equally spaced position angles. 
We inserted only one artificial planet at a time and ran our PSF subtraction for a range of PCA components $\in \left[5, 10, 20, 30, 40, 50, 75, 100\right]$. 
Based on these extensive artificial planet experiments, we computed the contrast curves based on a t-test \citep{Mawet2014}, assuming that the residual noise was sufficiently Gaussian distributed. 
We set the detection threshold to a false positive fraction (FPF) of $2.87 \times 10^{-7}$, which is equivalent to $5\sigma$ at large separations from the star.
The samples for the t-test were based on aperture sums as originally suggested by \cite{Mawet2014}. We accounted for the effect of the aperture placement by considering 20 different aperture positions for each test (see \citealt{Bonse2023}).
The final contrast curves reach out to 5 arcsec (the maximum apparent separation allowed by our data) and are shown in Fig. \ref{fig:HCI_cc_ml}. We further converted the obtained contrast curves to mass limits assuming the stellar ages in Table \ref{tab:star_par} and the AMES-Cond evolutionary models \citep{Baraffe2003}. For stars without literature values, we assumed a stellar age of $5\pm0.5$ Gyr. The resulting mass limits are also shown in Fig. \ref{fig:HCI_cc_ml}.

\subsection{Mass limits from radial velocity}\label{sec:RV_ml}

\begin{figure*}
\includegraphics[width=\textwidth]{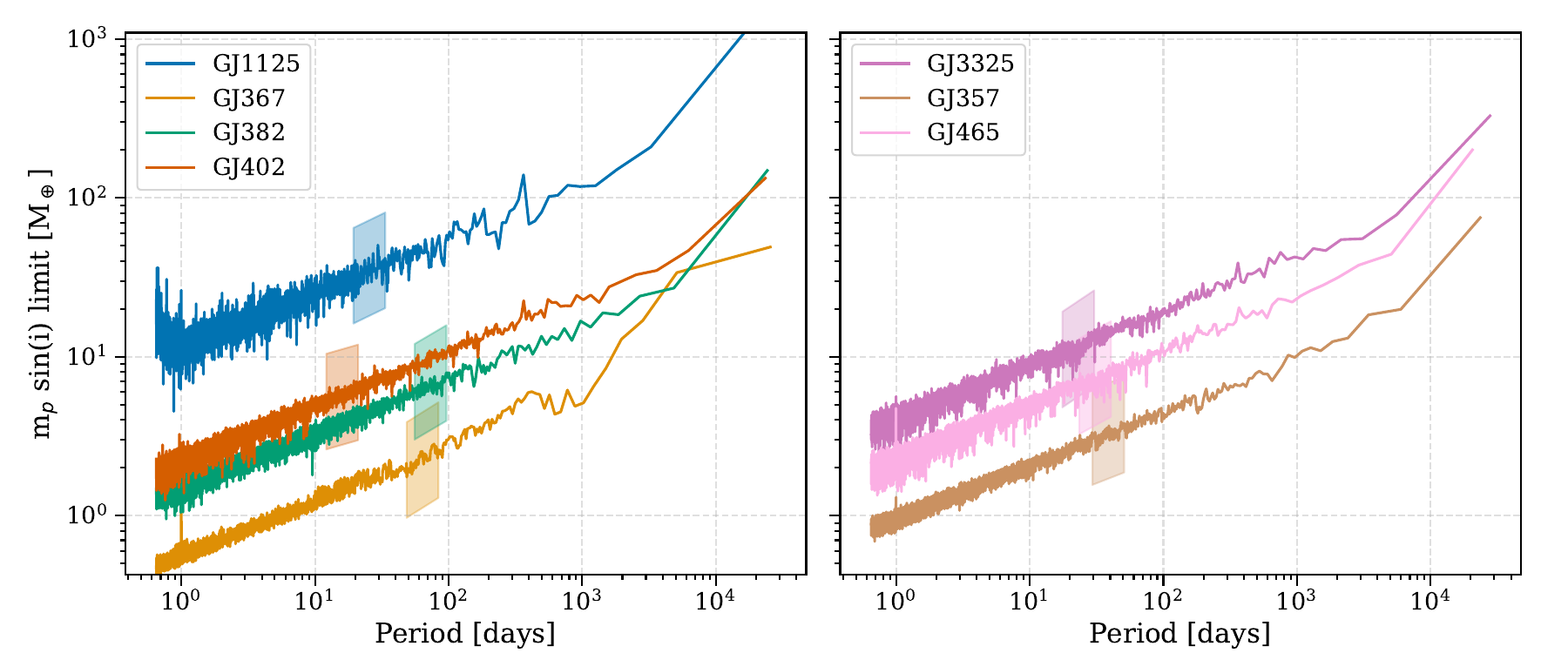}
\caption{Mass limits computed from the archival RV data using a bootstrap method (straight lines, following \protect{\citealt{Bonfils2013}} and references therein). The shadowed regions indicate the habitable zones (HZ, same color coding as for the mass limits). The targets have been split into two plots for ease of reading. The available RV data would allow us to detect planets in the HZ with minumum mass m$_p~\sin(i)~=~1-30~$M$_{\odot}$, depending on the target. Everything above the mass limits should have been detected. See Section \protect{\ref{sec:RV_ml}} for details.}
\label{fig:RV_ml}
\end{figure*}

As discussed in Section \ref{sec:RV_data}, apart from GJ~367 the archival RV data do not show any evidence of companions orbiting the target stars. We can however use the RV timeseries to compute the mass limits for potential planets around these stars, i.e. to constrain the companion mass / semi-major axis parameter space where potential planets could still be present, but whose RV signal would be too weak to be detected with the currently available data. The mass limits as a function of both orbital period $P$ and semi-major axis $a$ are shown in Fig. \ref{fig:RV_ml}. Because of geometry, the companion mass m$_p$ is degenerate with the companion's orbit inclination $i$, so that only a minimum mass m$_p \sin(i)$ can be computed.

To obtain the mass limits based on the RV observations we followed the procedure proposed in \cite{Bonfils2013}. In a first step, we created 1000 time series by randomly shuffling the original RV timeseries, computed the correspondent Lomb-Scargle periodograms as in Section \ref{sec:RV_data}, and found the maximum power of each periodogram, pow$_{\rm{max}, i}$. The power that divides the lower $99\%$ from the upper $1\%$ of the values in $\{$pow$_{\rm{max}, 1}, ...,  $pow$_{\rm{max}, 1000}\}$ represents the $1\%$ false alarm probability (FAP), which will be used in the following of the analysis. For each period $P_{\rm{GLS}}$ considered in the original Lomb-Scargle periodogram we then simulated 12 RV time series by adding to the original RV data a sinusoidal with period $P_{\rm{GLS}}$, an initial semi-amplitude $K_{\rm{in}} = 0.5$ m s$^{-1}$, and one of 12 equi-spaced phases $T$. Finally, for each period $P_{\rm{GLS}}$ and phase $T$ we increased $K_{\rm{in}}$ until the correspondent signal in the Lomb-Scargle periodogram reached the $1\%$ FAP defined above, and converted the obtained $K$ to m$_p \sin(i)$ assuming Keplerian orbits and the stellar mass listed in Table \ref{tab:star_par}. For each period $P_{\rm{GLS}}$ the mass limit corresponds to the average over the 12 trial phases $T$.

The obtained mass limits for the 7 target stars with available archival RV data are shown in Fig. \ref{fig:RV_ml}. In the plots we also highlight the periods correspondent to the habitable zone (HZ) which we computed using the Bolometric correction from \cite{Habets1981} and the scaling values for the inner and outer radius from \cite{Kasting1993}, \cite{Kasting1996} and \cite{Whitmire1996}. We note that the available data would  allow to detect planets in the HZ with minumum mass m$_p \sin(i) = 3-30$~M$_\oplus$, depending on the target.

\subsection{Mass constraints from astrometry}\label{sec:ast}

The PMa is defined as the difference between the long-term proper motion (PM) vector measured between two astrometric surveys, in our case Hipparcos and Gaia EDR3, and the instantaneous PM vector measured from one of the two catalogues.
This quantity provides an additional constraint on the mass and semi-major axis of potential companions orbiting our target stars, therefore influencing its motion around the common barycenter and the displacement of its photocenter.

We computed these constraints following the procedure described in \cite{Kervella2019, Kervella2022} and the PMa measurements in the associated catalogue (\citealt{Kervella2021}). Specifically, for the simplest case of m$_{p} \ll $ M$_{\star}$ and a circular orbit perpendicular to the line-of-sight, the relation between the companion mass m$_{p}$, the stellar mass M$_{\star}$ and the semi-major axis (orbital radius) $r$ is given by:

\begin{equation}
    \frac{{\rm{m}}_p}{\sqrt{r}} = \sqrt{\frac{{\rm{M}}_{\star}}{G}}v_{1} = \sqrt{\frac{{\rm{M}}_{\star}}{G}} \left( \frac{\Delta \mu \text{ [mas yr$^{-1}$]}}{\varpi \text{ [mas]}} \times 4740.47 \right)
\label{eq:sc}
\end{equation}

where $v_1$ is the tangential orbital velocity of the primary star, $G$ the Gravitational constant, $\Delta \mu$ the PMa and $\varpi$ the parallax (see Eq. 7 in \citealt{Kervella2019}). From Eq. \ref{eq:sc} we obtain the companion mass sensitivity curve m$_{p}(r) = f(r)$.

In a more realistic case, we need to take into account the uncertainty in the orbital inclination as well as observing window smearing and efficiency as a function of orbital period (see Section 3.6 in \citet{Kervella2019} for details about statistical and systematic corrections), so that the normalised companion mass sensitivity curve is given by:

\begin{equation}
    {\rm{m}}_p(r) = \frac{\sqrt{r}}{\gamma [ P(r) / \delta t_{{\rm{GDR3}}} ]} \sqrt{\frac{{\rm{M}}_{\star}}{G}} \frac{\Delta v_{{\rm{T, GDR3}}}}{\eta \zeta}
    \label{eq:sc_norm}
\end{equation}

where $P(r)$ is the period corresponding to the orbital radius $r$ (semi-major axis), $\Delta v_{{\rm{T, GDR3}}}$ is the norm of the tangential PMa vector converted to linear velocity using the Gaia DR3 parallax (from \citealt{Kervella2021}), $\gamma$ and $\zeta$ are two functions quantifying the window smearing and the efficiency as defined in \cite{Kervella2019}, $\eta = 87^{+12}_{-32}\%$ is a normalisation factor which accounts for the uncertainty in the orbital inclination (\citealt{Kervella2019}), and $\delta t_{\rm{GDR3}} = 1038$d is the observing window for the Gaia DR3 mission (\citealt{GaiaCollaboration2021}).

\begin{figure*}
\includegraphics[width=\textwidth]{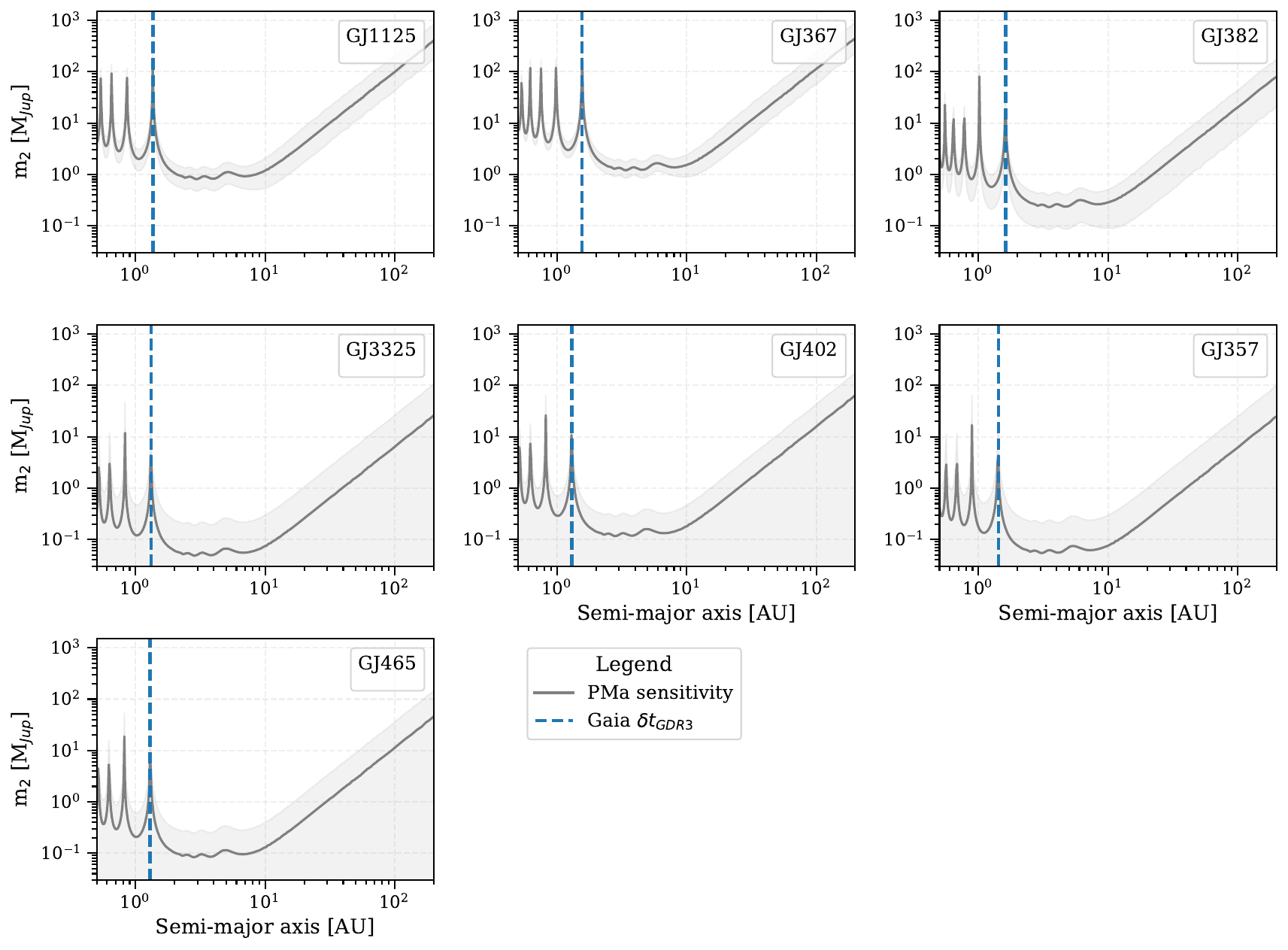}
\caption{PMa sensitivity limits (gray) computed for the seven targets present in both the Hipparcos and Gaia EDR3 catalogues (see Section \ref{sec:ast} for details). The shaded region corresponds to the $1\sigma$ uncertainty. The blue dashed line indicates the semi-major axis correspondent to the observing window for the Gaia DR3 mission $\delta t_{\rm{GDR3}} = 1038$d.}
\label{fig:astr_ml}
\end{figure*}

The PMa sensitivity limits, i.e. the constraints on companion mass m$_p$ and semi-major axis $a$, which we computed for the seven targets present in both the Hipparcos and Gaia EDR3 catalogues using Eq.~\ref{eq:sc_norm} and the PMa values in \citealt{Kervella2021}, are shown in Fig.~\ref{fig:astr_ml}. We note that these curves can be interpreted as a sort of mass limits. Indeed, the curves describe the possible $(m_p,r)$ combinations, given the known stellar mass $M_{\star}$ and the observed tangential orbital velocity $v_1$. By keeping $M_{\star}$ fixed, every planet with $(m_p,r)$ lying above the derived curve would have a tangential velocity $v > v_1$, and this would have been detected. Similarly, everything lying below would have $v < v_1$, and could not be distinguished/detected.

\subsection{Combining the constraints}\label{sec:combine}

\cite{Boehle2019} introduced a Monte Carlo method to estimate the fraction of companions that would be detected for a given companion mass and semi-major axis, based on constraints on planetary architecture from HCI and RV. In contrast to the mass constraints from RV described above, these calculations allow the direct constraint of the companion mass itself, $m_p$, rather than the minimum mass $m_p \sin(i)$. This approach also allows us to test for the presence of planets with periods much larger than the RV baseline (and thus larger semi-major axis), whose signal would not be seen as periodic in the available RV data and would therefore be barely detectable in the periodogram. Also, it allows us to test for the presence of planets with semi-major axes larger than the HCI field of view (because of projection effects), at distances where we cannot compute the contrast curves.

We expanded this approach to incorporate constraints from astrometry and applied it to the targets in our sample. 
First, for every semi-major axis and mass in a grid with $0 < a <$ 100 AU (25 points) and $0 <$ m$_p <$ 55 M$_{\rm{Jup}}$ (17 points) we generated $10'000$ Keplerian orbits by randomly drawing the inclination $i$ (for randomly oriented orbital planes) and the argument of periastron $\omega$ (uniform distribution spanning $0^{\circ}-360^{\circ}$), while keeping the eccentricity set to 0 (as in \citet{Boehle2019} we only consider the conservative case of circular orbits\footnote{This choice is justified as the targets in our sample are relatively old and therefore likely to host companions on dynamically evolved, low-eccentricity orbits. Furthermore, eccentric orbits can produce higher RV amplitudes, potentially enhancing detectability. }). For each orbit, we then computed the expected RV signal at the epochs of the RV observations, the projected separation at the epoch of the SPHERE HCI observation, and the norm of the tangential PMa vector converted to linear velocity using the Gaia DR3 parallax. For each observing technique, we then determined whether a companion on a given orbit would have been detected as follows: For RV, we considered the planet signal as detectable if the maximum RV difference measured across the simulated observations (i.e. RV$_{max}$ - RV$_{min}$) is higher than 5 times the standard deviation of the measured RVs\footnote{For the combined analysis, we follow the procedure and threshold outlined in \citealt{Boehle2019}, rather than the periodogram-based criterion presented in Section \ref{sec:RV_ml}, because the semi-major axes of interest correspond to orbital periods longer than the RV time baseline, making the corresponding RV signals non-periodic in the available RV time series.}. For HCI, the planet is considered as detectable if its mass is higher than the mass limits computed from the contrast curve at the predicted separation. Finally, for astrometry a planet is considered as detectable if the derived $\Delta v_{{\rm{T, GDR3}}}$, including the statistical and systematic corrections described in Section \ref{sec:ast}, is equal or higher than the one reported in \cite{Kervella2021}. The completeness for each combination of mass and semi-major axis was then determined from the percentage of detectable planets in the corresponding bin. 

\begin{figure*}
\includegraphics[width=\textwidth]{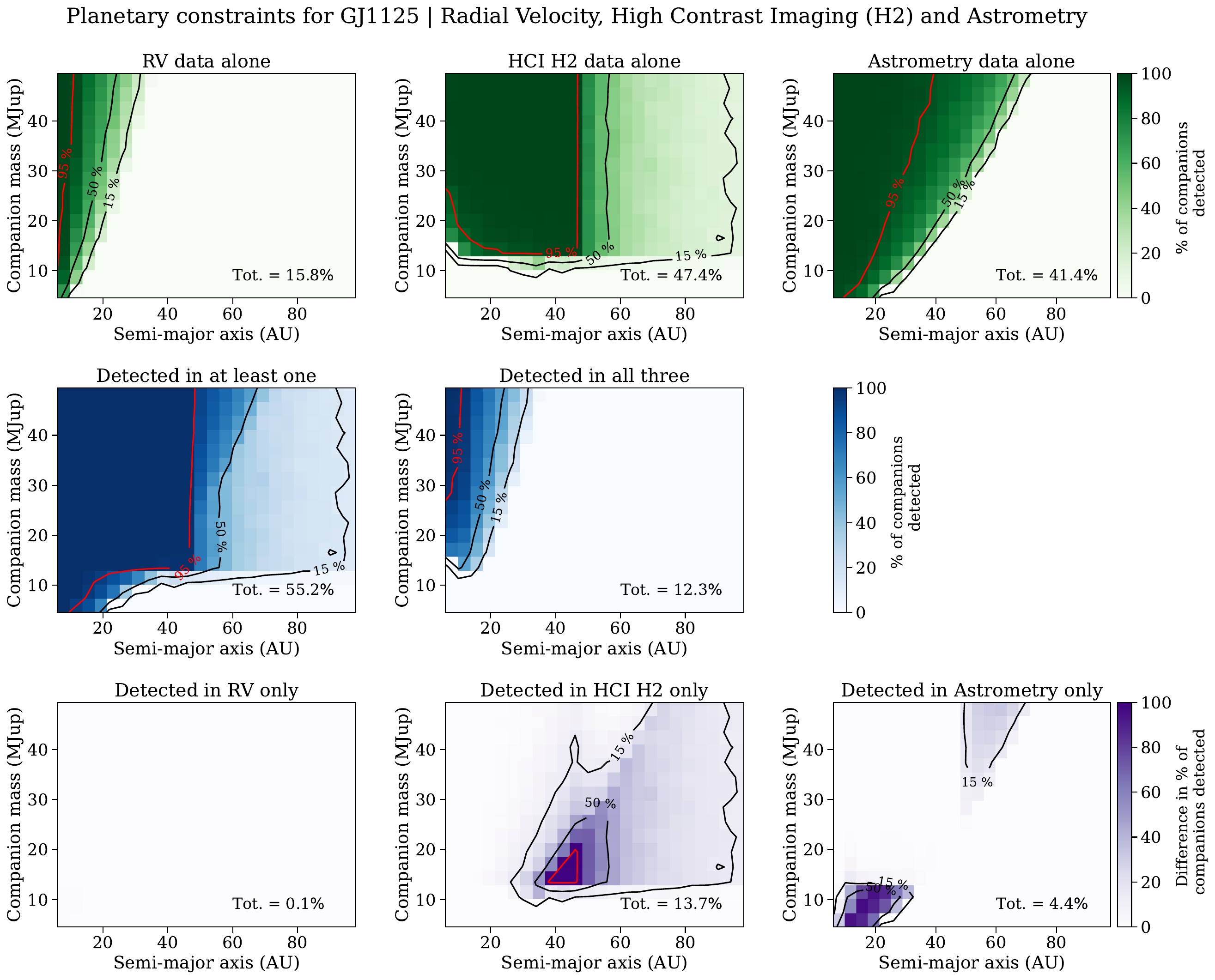}
\caption{Constraints on companion mass and semi-major axis for GJ1125 (example source, see Appendix \ref{app:com} for the whole sample). \textit{Top row (green)}: Percentage of companions that can be detected by radial velocity, SPHERE high contrast imaging (H2) and astrometry data alone. \textit{Middle row (blue)} Percentage of companions that can be detected by combining the three techniques (left), and that are detected by all the techniques (right). \textit{Bottom row (purple)}: Percentage of companions that can be detected exclusively by each technique. In each plot, `Tot' refers to the total detection ration integrated over the whole grid. See Sections \ref{sec:combine}, \ref{sec:const_ind} and \ref{sec:const_comb} for more details.}
\label{fig:comb_ex_GJ1125_H2}
\end{figure*}

\begin{figure*}
\includegraphics[width=\textwidth]{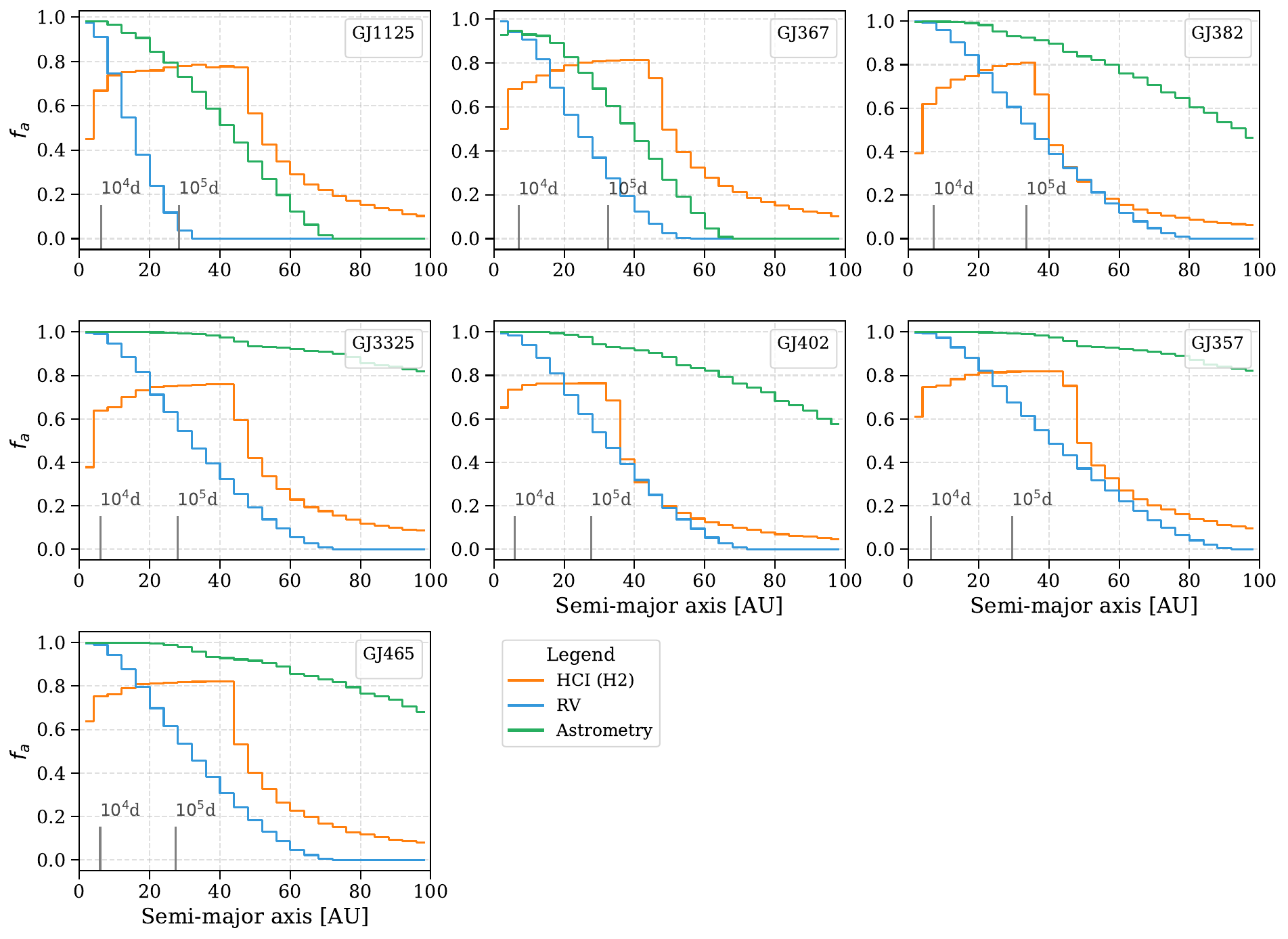}
\caption{Fraction of planets that could be detected as a function of semi-major axis by the three methods (HCI, RV and astrometry) separately. For each star we project the results obtained by the three methods separately (Fig. \ref{fig:comb_ex_GJ1125_H2} and Fig. \ref{fig:comb_ex_GJ367_H2} - \ref{fig:comb_ex_GJ357_H2}) on the x-axis and show the fraction of planets that could be detected as a function of semi-major axis.}
\label{fig:perc_a}
\end{figure*}

The resulting constraints for our target stars are shown in Fig. \ref{fig:comb_ex_GJ1125_H2} and Fig. \ref{fig:comb_ex_GJ367_H2} - \ref{fig:comb_ex_GJ357_H2}. For each star, the top row shows the percentage of companions (orbits) that could be detected by RV, SPHERE HCI (H2) and astrometry data alone. The middle row shows the percentage of companions that could be detected by combining the three techniques (left), and that are detected by all the techniques (right). Finally, the bottom row shows the percentage of companions that could be detected exclusively by each technique. In Fig. \ref{fig:perc_a}, for each star we project the results obtained by the three methods separately on the x-axis and show the fraction of planets that could be detected as a function of semi-major axis.

\section{Results}\label{sec:results}

\subsection{Contrast and mass limits from individual methods}\label{sec:const_ind}

Fig. \ref{fig:HCI_cc_ml} (1st and 2nd rows) shows the limiting contrast curves in the SPHERE \textit{H2} and \textit{H3} bands extending up to 5 arcsec, corresponding to (projected) separations of $25 - 50$~AU depending on the distance of the star. The limiting contrast for both filters goes down to $14-17$~mag in the background-limited region at angular separations $> 2$~arcsec. In Fig. \ref{fig:HCI_cc_ml} (3rd and 4th rows) we show the same contrast curves converted to mass limits using the AMES-Cond evolutionary models (\citealt{Baraffe2003}) for the given stellar ages, distances and apparent magnitudes. Our \textit{H2} images allow the detection of companions with masses down to $5-10$~M$_J$ in the background-limited regions, and a factor $\sim2$ higher in the \textit{H3} band. This difference is due to the fact that our observations probe the temperature regime $T < 1000$K where methane absorption is strong, so the flux in the methane absorption band is very weak (compared to the continuum). As the \textit{H2} filter provides better limits compared to \textit{H3}, we will concentrate on these observations only. We note that for two targets the stellar age is not known, so we assumed $5 \pm 0.5$~Gyr, consistent with other work in the PSION project. We have tested that for these targets, and for the other targets where the age is known, varying the age within the (assumed) uncertainties does not significantly change the results.

The mass limits from RV are shown in Fig. \ref{fig:RV_ml}. The archival observations allow us to compute limits up to $\sim 10^4$ days, corresponding to semi-major axis of $\sim 10$~AU. As expected, the limiting minimum mass m$_p \sin(i)$ shows a power-law increase with period. These observations would allow to detect planets in the HZ with minimum masses m$_p~\sin(i)~=~1-30~$M$_{\odot}$.

The companion mass sensitivity limit from PMa (astrometry) is shown in Fig. \ref{fig:astr_ml}, extending up to $\sim200$~AU. Unlike RV, these curves already take into account the unknown inclination $i$, so the y-axis represents the companion mass itself and not a minimum mass. Due to observational window smearing and efficiency, the sensitivity at separations below the Gaia DR3 observing window $\delta t_{\rm{GDR}} = 1038$d ($1-2$~AU depending on the source) shows strong peaks and is difficult to interpret (at least in the context of this study). At larger separation, the sensitivity flattens between $\sim 2-10$ AU reaching a minimum companion mass of $0.1 - 1$~M$_J$ depending on the source ($6 \times 10^{-2}$ for GJ 357), and shows a power-law increases at larger separations.

These results clearly show how the different detection methods are sensitive to very different parts of parameter space. This is also well illustrated in Fig. \ref{fig:perc_a}, which shows, for each target, the fraction of simulated planets that would be detected by each technique as a function of semi-major axis up to 100 AU. Specifically, RV and astrometry would detect the majority (up to $\sim 100\%$) of close-in planets, and the detection fraction decreases rapidly with increasing semi-major axis, although the decrease is less pronounced for astrometry. The RV detection fraction indeed drops to $\sim 40\%$ at $20-40$~AU for most sources, while the astrometry detection fraction remains above $\sim80\%$ out to $\sim 60$AU for 5 out of 7 targets. On the other hand, HCI misses up to $\sim60\%$ of the simulated companions at $a < 10$~AU. The detection rate then slowly reaches a $\sim 80\%$ ``plateau'' between $\sim 20-40$~AU, and drops steeply after that.

\subsection{Architecture constraints from combining methods}\label{sec:const_comb}

Fig. \ref{fig:comb_ex_GJ1125_H2} and Fig. \ref{fig:comb_ex_GJ367_H2} - \ref{fig:comb_ex_GJ357_H2} show the percentage of the 1000 simulated companions for each mass/semi-major axis combination that could be detected by RV, HCI and astrometry, and by combinations of these methods. We are considering semi-major axes up to 2 times the SPHERE FOV, as companions orbiting at those separations can still have projected separations within the FOV for some orbital phases. Similar to what we noted for Fig. \ref{fig:perc_a}, the limiting mass for RV and astrometry increases quickly as a function of semi-major axis, whereas the constraints from HCI are U-shaped with a flat plateau corresponding to the background limited-region and a sharp decrease at shorter separations as well as beyond the SPHERE FOV.

Looking at the methods individually, for 5 of the 7 targets the best performance is provided by astrometry, which would be able to detect up to $\sim81-94\%$ of the simulated companions. For GJ~1125 and GJ~367, the best performance is provided by HCI, but with a lower percentage of detectable companions ($\sim40\%$). The worst performance is always provided by RV, with $\sim15\%$ of detectable companions for the most distant star GJ~1125, and $\sim35-44\%$ of detectable companions for the other targets. This result is not surprising as RV is mostly sensitive to short period companions and is therefore not optimal at the largest semi-major axes considered in this study (see also target selection in Section \ref{sec:sample_sel}).

Combining all three methods, the percentage of detectable targets ranges from $\sim50\%$ to $\sim94\%$ among the 7 targets. Again, in 5 cases, astrometry provides the greatest enhancement compared to the detection fraction reached by the other two techniques, with $\sim34-43\%$ of companions detected exclusively by this technique. These companions usually lie in the lowest mass / shortest semi-major axis or highest mass / middle to largest semi-major axis parts of the considered parameter space, as these are not accessible to the other two techniques. For GJ~1125 and GJ~367 the greatest enhancement, is provided by HCI, with $\sim9-14\%$ of the simulated companions detected only by this technique, in agreement with \citealt{Boehle2019}. These companions lie in a triangular shape, bounded at the bottom by the HCI mass limits in the background-limited regime and at the side by the RV and/or astrometry mass limits.

\subsection{Survey sensitivity and companion population modeling}\label{sec:sd}

Despite our combined analysis allowing us to search a rather large parameter space, we did not detect any new companion. This is not surprising given the known statistics of large planets and brown dwarfs. As discussed in \cite{Boehle2019}, HCI studies show that giant planets on a wide orbit are quite rare. For example, \cite{Bowler2016} analysed a sample of FGKM stars and found that $<3.9\%$ of M dwarfs host giant planets. Similarly, \cite{Meyer2018} applied combined statistic on a sample of M stars and obtained a planet detection rate of $1.5-3\%$ in the parameter range $10 < a <$ 1000 AU and $1 <$ m$_p <$ 10 M$_{\rm{Jup}}$. 
This means that, at least statistically, we would expect $<1$ giant companion among our 7 targets in the considered parameter space.

\begin{figure*}
\includegraphics[width=\textwidth]{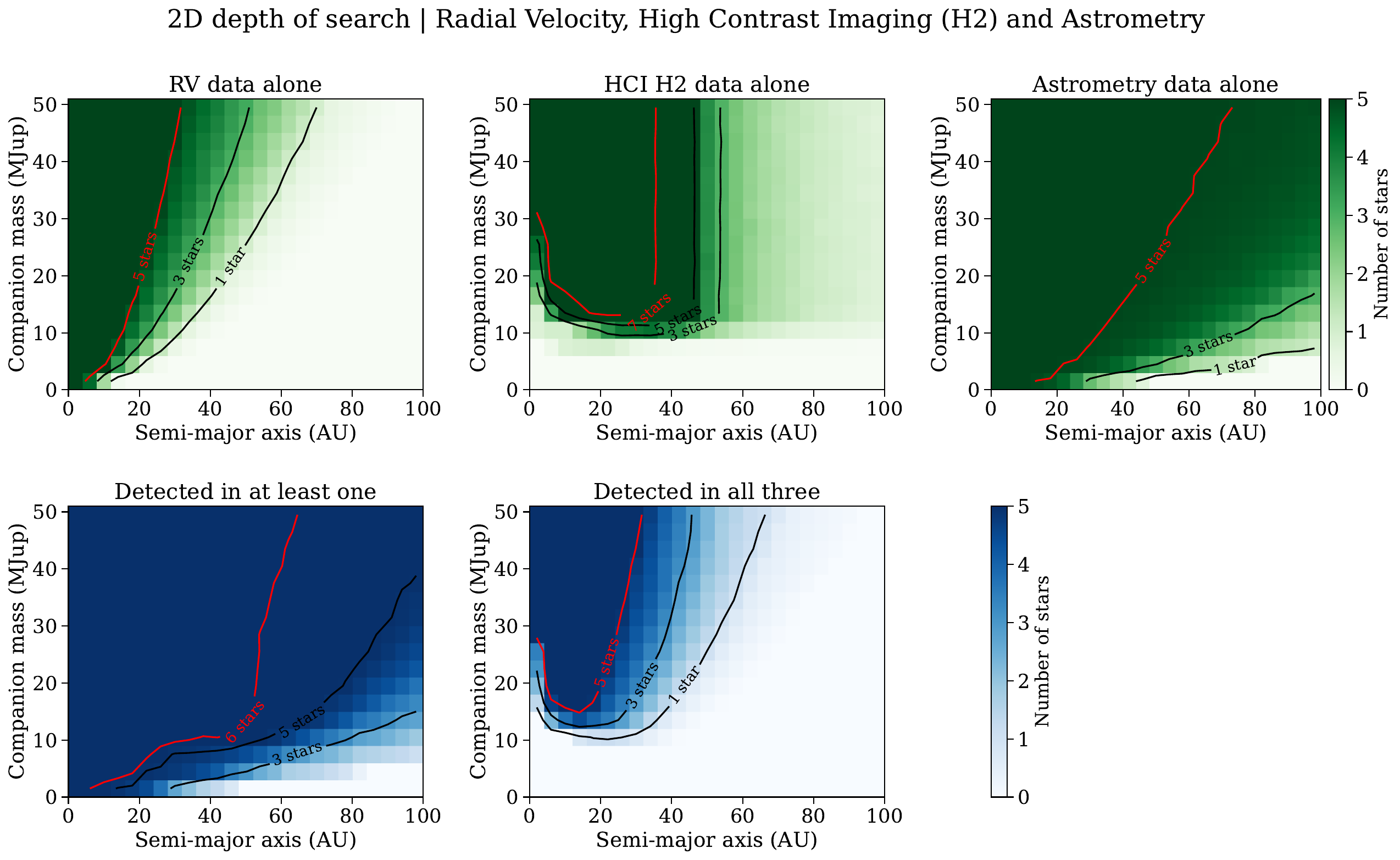}
\caption{Depth of search for the three detection methods separately (top) and combined (bottom). The 2D survey depth (or depth of search of the survey) gives the number of stars around which the survey is sensitive for a given companion mass and semi-major axis. See Section \ref{sec:sd} for details.}
\label{fig:survey_depth}
\end{figure*}

\begin{figure*}
\includegraphics[width=\textwidth]{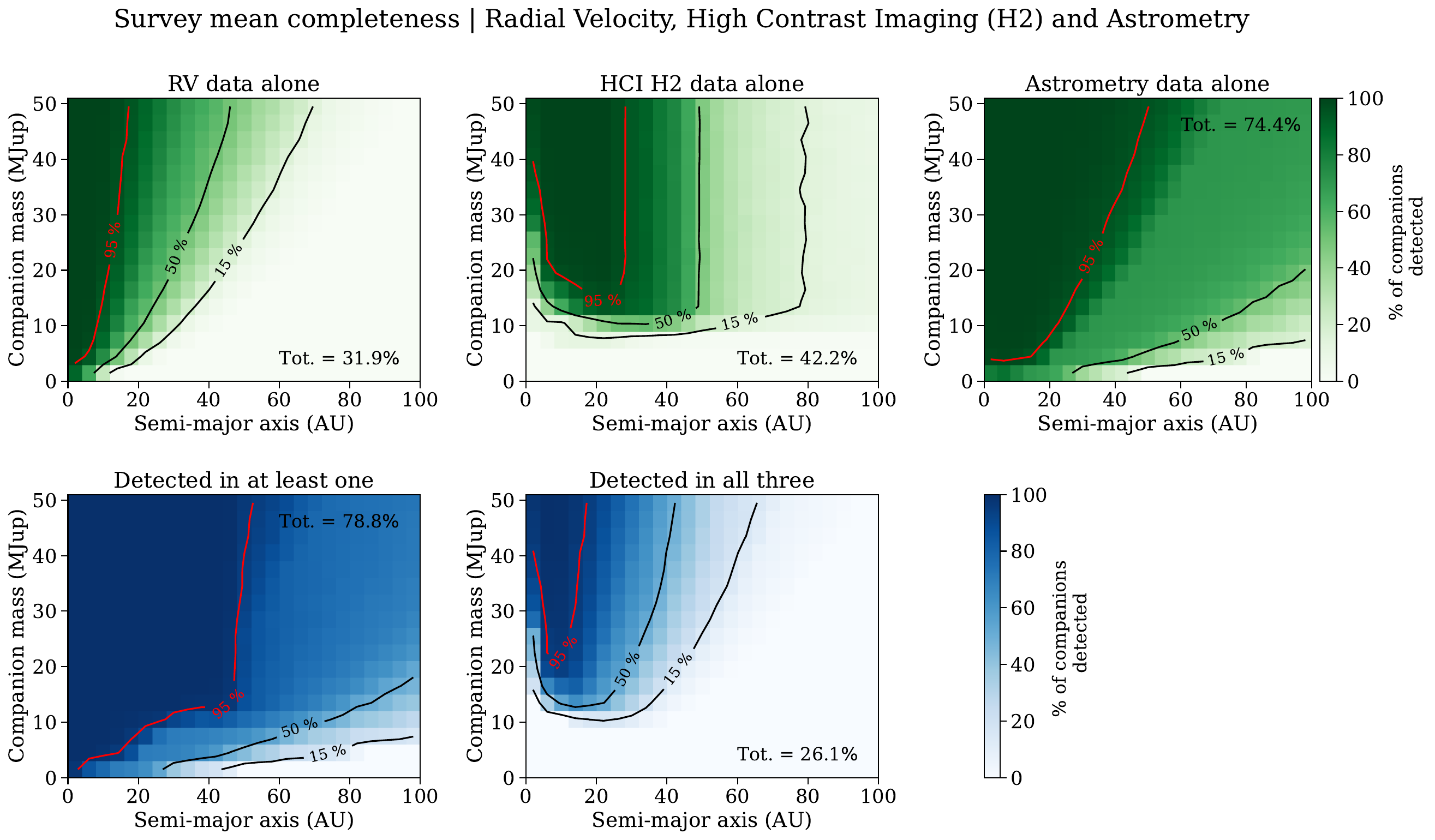}
\caption{Survey mean completeness for the three detection methods separately (top) and combined (bottom). The 2D survey (mean) completeness is computed as the average of the detection maps obtained for the individual targets. See Section \ref{sec:sd} for details.}
\label{fig:survey_comp}
\end{figure*}

Although 7 targets cannot be considered a sample in statistical sense, we still decided to follow \cite{Vigan2021} to quantitatively estimate the number of expected detections from our analysis by computing the 2D survey depth and 2D survey completeness of our ``sample'' and comparing it to population models. Specifically, the 2D survey depth (or depth of search of the survey, Fig. \ref{fig:survey_depth}) gives the number of stars around which the survey is sensitive for a given companion mass and semi-major axis. This is computed as the sum of the detection maps obtained in Section \ref{sec:combine}. For the combined analysis (``detected in at least one method" in Fig. \ref{fig:survey_depth}), the core of the sensitivity ($>6$ stars) reaches $50-60$ AU for companions with masses $> 10$M$_{\rm{Jup}}$. Similarly, the 2D survey (mean) completeness (Fig. \ref{fig:survey_comp}) is the average of the individual detection maps. Again for the combined analysis the sensitivity is $>95\%$ for companions with masses $> 10$M$_{\rm{Jup}}$ up to separations $a = 50-60$ AU.

To compare our survey results with theoretical expectations, we adopt a phenomenological model based on Meyer et al. (in prep.). This parametric model simultaneously fits the frequencies of both very low-mass brown dwarfs formed through ``multiple-like processes" (fragmentation during early cloud collapse or disk fragmentation analogous to binary or multiple star formation), and gas giants formed via ``planet-like processes", such as core accretion or disk instability within a protoplanetary disk. These frequencies are modelled as functions of orbital separation and host star mass. The frequency data is derived from 50 independent point estimates compiled over 20 years of direct imaging, microlensing, and radial velocity surveys (e.g. \citealt{Wittenmyer2016},  \citealt{Borgniet2017}, \citealt{Nielsen2019}, \citealt{Vigan2021}, \citealt{Fulton2021}, \citealt{Grandjean2022}). The parameter space encompasses companion masses ($m_{c} \sim 1-85$ M$_{J}$), orbital separations ($a \sim 1-6000$ AU), and stellar masses ($M_{*} \sim 0.2-3.0$ M$_{\odot}$). We assume log-normal orbital separations and power-law mass ratio distributions for both planets and brown dwarfs around M dwarfs. The brown dwarf orbital distribution is fixed to the values in \citealt{Winters2019} ($\mu_{BD}=1.43$, $\sigma_{BD}=1.21$) , with the projected separation (20 AU) adjusted to the physical separation (27 AU) using a correction factor of 1.35 from \cite{Duquennoy1991}. The final model is characterized by six parameters: two normalization factors for giant planets (GP) and brown dwarfs (BD) ($A_{GP}$, $A_{BD}$), two parameters for the log-normal semi-major axis distribution of gas giants ($\mu_{GP}$, $\sigma_{GP}$), and two power-law indices for the mass ratio distributions of giant planets and brown dwarfs ($\alpha_{GP}$, $\beta_{BD}$). We found ln$(A_{GP}) = -5.52_{-0.83}^{+0.93}$, ln$(\sigma_{GP}) = 0.53_{-0.07}^{+0.06}$, ln$(\mu_{GP}) = -1.32_{-0.22}^{+0.18}$, $\alpha_{GP} = 1.43_{-0.13}^{+0.12}$, ln$(A_{BD}) = -3.78_{-0.47}^{+0.70}$, and $\beta_{BD} = 0.36_{-0.34}^{+0.30}$. These results show that giant planets peak at orbital distances of 4 AU and taper off beyond the 1-10 AU range, consistent with the latest exoplanet survey findings (e.g. \citealt{Cumming2008}, \citealt{Fulton2021}, \citealt{Meyer2018}).

For our program, we model companions around M dwarfs as two distinct populations: (1) Giant Planets (0.03M$_{J}$–0.1 M$_{\odot}$), and (2) Brown Dwarfs (3–75 M$_{J}$). The lower mass limit for giant planets is set at 0.03 M$_{J}$ (or 10 Earth masses), while the upper limit is defined as 10\% of the stellar mass (0.1 M$_{\odot}$), rather than the conventional deuterium-burning threshold at 13 M$_{J}$, to better reflect the maximum mass typically expected from disk-based formation processes. For planetary masses below 1 M$_{J}$, we employ a custom sub-population model to describe so-called sub-Jovian planets, motivated by both the architecture of the Solar System and recent mm-wave ALMA disk continuum observations that show evidence of disk gaps possibly caused by embedded planets less than a few 10$M_{\oplus}$ (\citealt{Boekel2017}). For brown dwarfs, the lower mass limit is determined by the opacity limit for disk fragmentation, while the upper bound is set by the hydrogen-burning threshold of 75~M$_{J}$.

\begin{figure*}
\includegraphics[width=\textwidth]{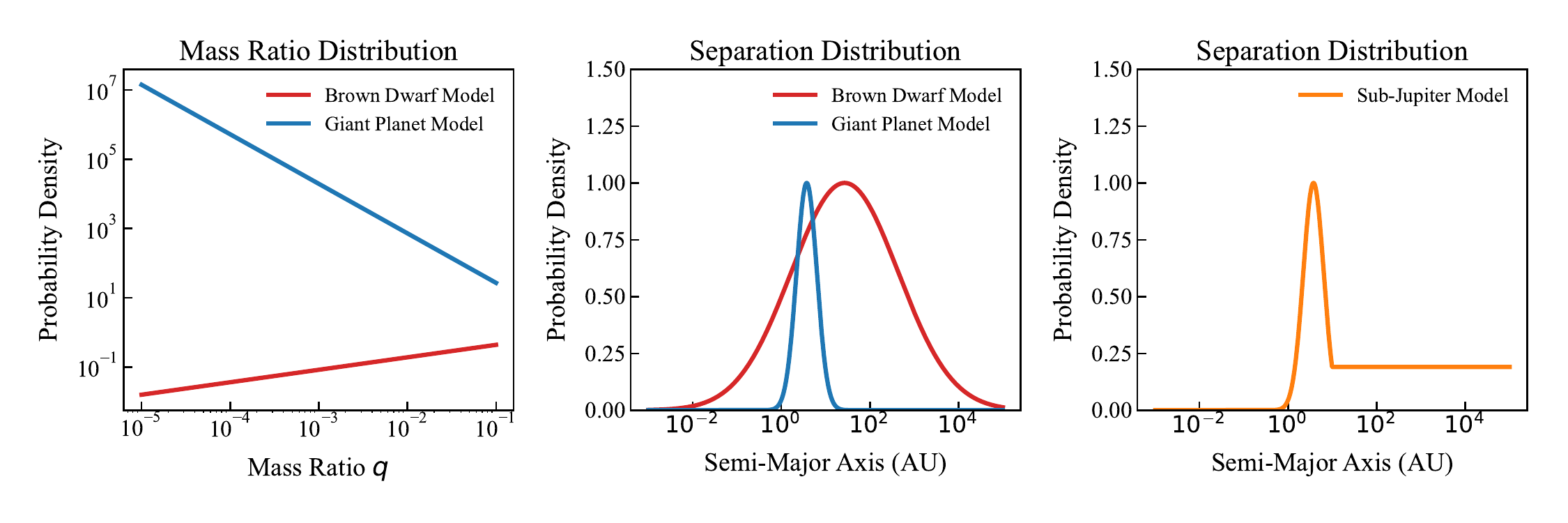}
\caption{Frequency distribution for models. The semi-major axis distribution for generated companions follow a lognormal distribution for the giant planet model from Meyer et al. in prep., a lognormal joint with a log-uniform distribution beyond 10 AU for the sub-Jupiter model, and the brown dwarf model uses \cite{Winters2019}, corrected to physical separation. Here, the sub-Jupiter model is a customized model where we have assumed the frequency within 10 AU is double that of the super-Jupiter/gas giant population, but log-flat beyond 10 AU. We normalise the distributions according to Meyer et al. in prep. and integrate over the mass-orbital space to estimate companion frequencies out to a certain radius.}
\label{fig:img_freq}
\end{figure*}

The mass/separation distribution models are defined as follows. For giant planets and brown dwarfs, we adopt the log-normal orbital distributions (Fig. \ref{fig:img_freq}) and power-law mass ratio distributions from the Meyer et al. (in prep.) model. For the sub-Jupiter mass planets, we used the same power-law distribution for the masses, extrapolating it from the super-Jupiter ($> 1 M_{J}$) regime into the sub-Jupiter ($< 1 M_{J}$) regime. For the separation distribution, we use a model where we double the super-Jupiter occurrence rate within 10 AU but maintain a flat distribution in $\log(a)$ to the outer cutoff radius (orange curve in Fig. \ref{fig:img_freq}). This is because previous work by \cite{Fulton2021} showed that the occurrence rate of sub-Jovian planets (0.1 - 1 M$_{J}$) is enhanced by almost twice compared to that of the super-Jupiters for orbits between 1 and 10 AU, but unknown beyond 10 AU range. While this survey may not capture this sub $1 M_{J}$ discovery space, large-scale JWST programs, like \citet{Carter2024}, will shed light on this population of sub-Jupiter and sub-Saturn mass exoplanets beyond 10 AU.

We derive the companion occurrence rate by integrating over the mass-orbital space over the range of masses defined above and separations from 0 out to an outer cutoff radius constrained by the SPHERE FOV. The estimated total frequencies of sub-Jupiters, super-Jupiters, and brown dwarfs within the FOV are 10.6$\%$, 5.6$\%$, and 12.4$\%$, respectively. These occurrence rates are consistent with estimates from the literature (e.g., \citealt{Meyer2018}; \citealt{Susemiehl2022}). For companion generation, we created $10^5$ random sets of orbital parameters using Poisson statistics and the following priors: uniform priors for the longitude of the ascending node and the longitude of periastron, cosine priors for inclination, and Gaussian priors for eccentricity ($\mu$ = 0, $\sigma$ = 0.3, \citealt{Hogg2010}). For each set of orbital parameters, we simulated 360 positions evenly spaced in time and assessed detectability across an even grid of companion masses and semimajor axes. The same set of orbital parameters was applied to every target. For low-mass planets, we utilized the BEX evolutionary models (\citealt{Linder2019}), while for higher-mass companions, we used the ATMO 2020 models (\citealt{Phillips2020}).

\begin{figure}
\includegraphics[width=0.5\textwidth]{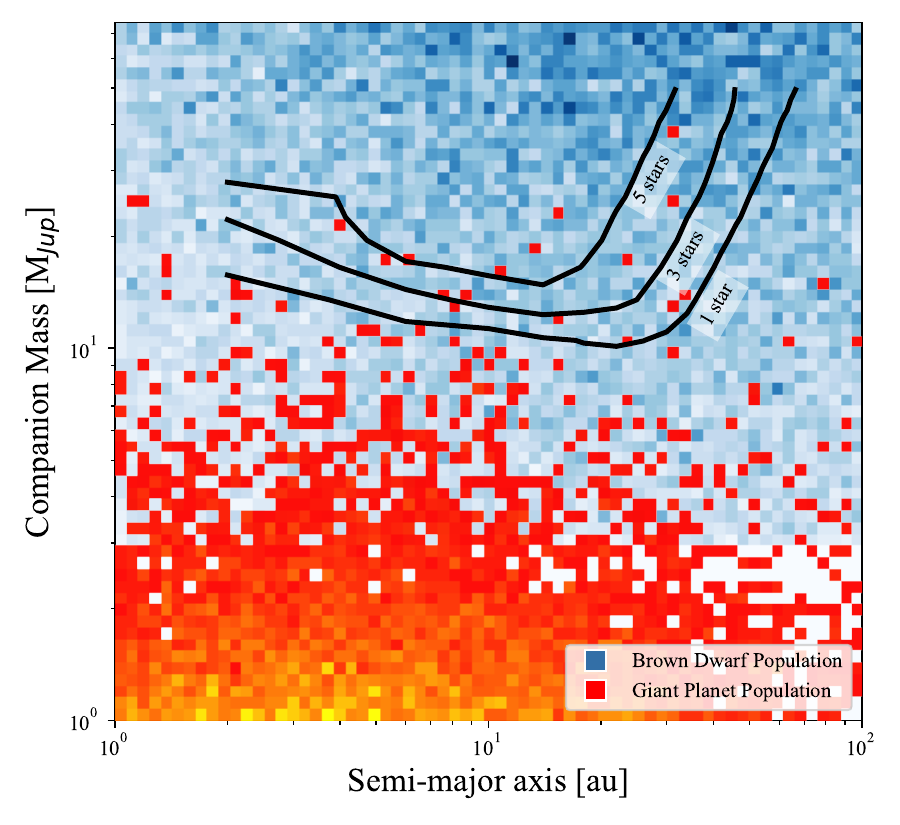}
\caption{Comparison of the depth of search of the SPHERE H2 survey for the presented 7 M dwarf sample (Fig. \ref{fig:survey_depth}) with a population of 100'000 draws from our parametric model for companions around M dwarfs. The contour lines give the numbers of stars around which the survey is sensitive to substellar companions as a function of mass and semimajor axis. The brown dwarf popluation generated from the model is represented with shades of blue (high density) to white (low density), while the giant planet part of the model is represented as a heatmap from red (low density) to yellow (high density of companions). }
\label{fig:model_mass_sep}
\end{figure}

The results from the modeling are illustrated in Fig. \ref{fig:model_mass_sep}, which shows the simulated populations of brown dwarfs and giant planets around M dwarfs against the SPHERE contrast curves. Based on the parametrisation described above, the code predicts that our HCI observations would detect no Sub-Jupiters, $\sim$0.01 Giant Planets and $\sim$0.26 Brown Dwarfs in our sample. This is consistent with our non-detections.

\section{Discussion and outlook}\label{sec:dis}

In this study, we have demonstrated that combining RV, HCI, and astrometric observations significantly enhances the constraints on the parameter space for potential companions, compared to the results obtained with any individual technique. We have quantified this improvement and found that the percentage of detectable companions within the considered parameter space ($a<100$ AU, m$_{\rm{c}} <50$ M$_J$) increases by up to $\sim 60\%$ compared to RV alone (improvement achieved for GJ~3325), $\sim 50\%$ compared to HCI alone (for GJ~3325), and $\sim 12\%$ compared to astrometry alone (for GJ~367). 
The relatively lower performance of RV is not surprising, as the parameter space considered in this study, both in terms of extension and resolution, has been optimised for the SPHERE observations, whereas RV techniques are generally more sensitive to companions at smaller semi-major axes. In contrast to \citet{Boehle2019}, we found no significant correlation between the magnitude of improvement and specific target or observational properties, such as stellar distance or the RV time baseline. A similar study focusing on a different region of parameter space, specifically in the planetary regime with shorter orbital periods and lower companion masses, using RV and astrometric data only, is currently ongoing (Sartori, Di Lenia et al., in prep).

In the current study we assumed that every point in the parameter space, i.e. every semi-major axis / companion mass pair, is in principle equally likely. Specifically, we did not consider any prior based on current statistics of known detections (e.g., from \citealt{Bowler2016}; \citealt{Meyer2018}; \citealt{Vigan2021}; \citealt{Bergsten2022}). In this way, we do not have inherited biases due to the sample selection and observational design of those studies. We also did not consider any potential information about the systems architecture from previous observations of our targets. For example, \citealt{Goffo2023} derived an inclination $i_{\rm b} = 79.89 \pm 0.85 \deg$ based on the TESS transit light curves of GJ~367. For this specific source, fixing the inclination to this high value would increase the percentage of possible detections with RV and astrometry ($+16.6\%$ and $+8.2\%$, respectively) and decrease the percentage for HCI ($-13.1\%$). The systematic inclusion of this type of prior information could be considered in the future as a further refinement of our method.

The presented method and results rely on already available data, specifically from observations which are (mostly) non-detections. Even in the now dawning ``era of big data''\footnote{New and upcoming telescopes and surveys such as the Vera Rubin Observatory (\citealt{Ivezic2008}; \citealt{LSST2009}, formerly known as LSST) and the Square Kilometer Array (SKA, \citealt{Braun2019}) will provide hundreds of PB of data, unprecedented in the history of astronomy.}, a re-analysis of archival observations with new methods and post-processing techniques is very valuable to extract important information and statistic about planetary and (sub-)stellar architecture even from data which were previously discarded. When planning new observations to search for companions, however, it is important to consider not only the parameter space explored (as it can be done with our method), but also the accessibility and cost of the data in terms of availability of telescopes and surveys, as well as the required observing time. While the PMa has to be computed based on two astrometric surveys (e.g., \citealt{Brandt2018,Brandt2019}; \citealt{Kervella2019,Kervella2019a,Kervella2021,Kervella2022}), in this case Hipparcos and Gaia, and is therefore only available for targets in those catalogues, RV and HCI observations can be obtained for many targets from currently operating telescopes. The sensitivity of RV strongly depends on its time baseline (e.g., \citealt{Sartori2023}), and requires multiple observations over years or even decades in order to confidently constrain the period and (minimum) mass of potential companions orbiting at large distances from the star. On the other hand, the constrast reached by HCI depends, among other factors, on the exposure time and on the amount of FOV rotation\footnote{The FOV rotation is crucial for the applicability of post-processing techniques such as ADI, e.g., \citealt{Marois2006}} (e.g., \citealt{Follette2023}). From the data analysis point of view, while HCI requires advanced post-processing techniques in order to robustly compute the contrast curves and confirm potential detections (e.g., \citealt{AmaraQuanz2012}; \citealt{Stolker2019}; \citealt{Bonse2023,Bonse2024}), RV is more affected by stellar variability, which is often difficult to disentangle from the companion signal (e.g., \citealt{Meunier2021}).

While SPHERE is one of the best instruments for HCI to date (e.g., \citealt{Follette2023}), future studies using e.g., the Enhanced Resolution Imager and Spectrograph (ERIS, \citealt{Davies2018}) at the VLT and the \textit{James Webb Space Telescope} (\textit{JWST}; \citealt{Gardner2006}) would offer significant improvements in both resolution and sensitivity. ERIS, with its advanced near- and mid-IR AO enabled imager NIX (\citealt{Parker2016}; \citealt{Boehle2021}), will improve the contrast performance at small angular separations. Notably, observations in the L and M bands (3-5 $\mu$m) will be particularly effective in detecting cooler, lower-mass companions such as giant planets and brown dwarfs. Additionally, the \textit{JWST} Near-InfraRed Camera (NIRCam, \citealt{Rieke2005}; \citealt{Rigby2023}), which has a similar wavelength coverage and angular resolution as NIX, and the Mid-Infrared Instrument (MIRI, \citealt{Glasse2015}; \citealt{Wright2023}) will offer an unprecedented sensitivity in the near- and mid-IR (0.6–28 $\mu$m), enabling the detection of even cooler and more distant companions. The combination of ERIS ground-based high-angular-resolution imaging and \textit{JWST} space-based sensitivity will significantly extend the parameter space for detecting and characterizing substellar companions, especially in cases where SPHERE H-band data might have missed fainter or cooler objects due to contrast limitations at smaller separations or at longer wavelengths. Moreover, combining archival HCI observations with new observations will provide a longer temporal baseline, allowing us to follow the orbital motion of (potential) companions between epochs, and thus to detect companions that were not visible in one of the observations because of their current position relative to the star. This is particularly important for nearby stars, such as those in our sample, where companions exhibit (apparent) larger orbital motions on short timescales, leading to improved detection completeness and better constraints on orbital parameters.

The RV measurements used in this study are from archival HIRES and HARPS observations, with time baselines ranging from 4090 days (GJ~1125) to 6961 days (GJ~332). In future studies, newer RV instruments such as the Echelle SPectrograph for Rocky Exoplanets and Stable Spectroscopic Observations (ESPRESSO, \citealt{Pepe2021}; \citealt{GonzalezHernandez2018}) at the VLT, and other upcoming high-precision spectrographs, will provide significant improvements in sensitivity compared to these archival data. ESPRESSO, with its ability to achieve an RV precision down to the cm s$^{-1}$ level, is particularly suited for detecting lower-mass companions, including super-Earths and giant planets in longer-period orbits, which may have been undetectable with previous instruments due to RV jitter and instrumental limitations. In addition, other instruments like EXtreme PREcision Spectrograph (EXPRES, \citealt{Jurgenson2016}) and the Near Infra Red Planet Searcher (NIRPS, \citealt{Bouchy2017}; \citealt{Wildi2017}), which operate in the optical and near-infrared regimes, respectively, will complement ESPRESSO by extending RV coverage to different wavelength ranges, reducing the impact of stellar activity on RV measurements, and improving sensitivity to cooler and low-mass companions around M-dwarfs. The increased precision, broader wavelength coverage, and improved stability offered by these next-generation RV instruments will allow us to probe a wider range of parameter space for detecting substellar companions, particularly for nearby stars, where the RV signal is stronger due to proximity. The combination of archival data with new high-precision RV measurements will also allow us to constrain long-term trends more robustly and identify companions with longer orbital periods that may not have been detectable in previous RV datasets.

Combining observations obtained with different techniques is crucial for population statistics, as it allows for the exploration of a wider parameter space and helps mitigate observational biases inherent to each method. This multi-technique approach leads to a more complete understanding of stellar-companion architectures and demographics across a broad range of parameters, such as semi-major axis, companion mass, and host star spectral type. In this study, we focus on the combination of RV, HCI, and astrometry, building on the work of \cite{Boehle2019}, who analyzed a similar sample but considered only RV and HCI. Nearby stars are however also prime targets for large monitoring surveys such as \textit{Kepler} (\citealt{Borucki2010}) and the Transiting Exoplanet Survey Satellite (\textit{TESS}; \citealt{Ricker2015}), which have the sensitivity to detect exoplanet transits and transit timing variations (TTV, e.g., \citealt{Agol2005}) for thousand of stars on the whole sky. These surveys complement RV and HCI data by probing shorter-period companions, especially for bright, nearby stars. Additionally, microlensing surveys provide valuable constraints on intermediate-separation and low-mass planets around low-mass stars, primarily M-dwarfs (e.g., \citealt{Gould2018}; \citealt{Meyer2018}). Incorporating data from such complementary techniques will further expand the parameter space covered by our combined approach, yielding a more robust understanding of planetary system architectures.

As we elaborated in Section \ref{sec:sd}, the fact that we did not detect any new companion is not surprising given the current statistic of known large planets and brown dwarfs (e.g., \citealt{Bowler2016}; \citealt{Meyer2018}; \citealt{Vigan2021}), which shows that large planets on wide orbits, as probed by this study, are quite rare. Indeed, based on such statistics, $<1$ companion is expected for our sample. These results are also consistent with the parametric model, which predicts no Sub-Jupiters, $\sim$0.01 Giant Planets and $\sim$0.26 Brown Dwarfs in our sample in the parameter space probed by the HCI data. As we mentioned above, as it includes only 7 targets, our sample cannot be considered a statistical sample yet. As part of the PSION project we are however working on increasing the number of sources for which our method can be applied. Once the sample is big enough, it will be interesting to compare the results, in particular the depth of search and the 2D sensitivity for the whole sample, to \textit{population} synthesis models based on both gravitational instability (e.g., \citealt{Boss1998}; \citealt{Vorobyov2013}) and core accretion (e.g., \citealt{Goldreich1973}; \citealt{Pollack1996}) formation mechanisms. As it has been done in the past for different samples (e.g., \citealt{Janson2011}; \citealt{Mordasini2017}; \citealt{Vigan2021}), comparing such models to observations will help discriminating and better understanding the formation process of giant planets and brown dwarfs. As our method allows to directly compare the parameter space probed by the different detection techniques, this comparison will also help understanging on which technique we should concentrate for future observations in order to better constrain such formation mechanisms %

In summary, this study demonstrates the effectiveness of multi-method approaches in constraining companion populations, and highlights the still unexploited potential of archival data that may have previously been overlooked due to non-detection. Although the small sample size limits the statistical significance of the results presented in this work, the methodology can be easily extended to larger samples and different regions of the parameter space. In this study, we focus on giant planets and brown dwarfs, for which constraining the parameter space is essential for understanding their formation and evolution. The identification of outer giants (like Jupiter in the Solar System) may be essential for the habitability of inner terrestrial planets, e.g. through early water supply by scattering of volatile-rich solids, and later as a safeguard against bombardment (e.g. \citealt{Sanchez2018}, \citealt{Zain2018}). Proving the absence of giants near the HZ is also important, as giants near the HZ would make orbits inside the HZ unstable.  In the future, applying a similar approach in the planetary regime, particularly for rocky exoplanets, will allow us to better constrain the parameter space for Earth analogues and other terrestrial planets. This in turn will inform the target lists for future space missions such as the Large Interferometer for Exoplanets (LIFE) and the Habitable Worlds Observatory (HWO), which aim to identify potentially habitable or even inhabited planets around nearby stars

\begin{acknowledgements}
      We thank the anonymous referee whose comments helped to improve the quality and readability of this paper. Based on observations collected at the European Southern Observatory under ESO programme 0104.C$-$0336. This work has been carried out within the framework of the National Centre of Competence in Research PlanetS supported by the Swiss National Science Foundation under grants 51NF40$\_$182901 and 51NF40$\_$205606. The authors acknowledge the financial support of the SNSF. L.F.S. thanks Gabriele Cugno for helpful discussions on high-contrast imaging.
\end{acknowledgements}

\noindent
\footnotesize{{\it{Authors contributions.}} L.F.S.: Conceptualization, Methodology, Software, Validation, Formal
Analysis, Investigation, Data Curation, Writing -- Original Draft, Visualisation; S.P.Q.:  Conceptualisation, Resources, Writing -- Review and Editing, Funding Acquisition; M.J.B.:
Software (\texttt{applefy}), Writing -- Original Draft, Review and Editing; Y.L.: Formal Analysis (parametric model),
Software, Writing -- Original Draft, Review and Editing; F.A.D.: Writing -- Review and Editing; C.L.: Conceptualisation, Writing -- Review and Editing; A.B.: Conceptualisation, Methodology, Data Acquisition.\\}

\noindent
\footnotesize{{\it{Software.}} This research made use of: Astropy\footnote{\url{https://www.astropy.org/}}, a community-developed core Python package for Astronomy (\citealt{Astropy2013,Astropy2018}); Matplotlib\footnote{\url{https://matplotlib.org/}} (\citealt{Hunter2007}); \texttt{PynPoint}\footnote{\url{https://pynpoint.readthedocs.io/en/latest/}} (\citealt{AmaraQuanz2012}, \citealt{Stolker2019});  \texttt{species} (\citealt{Stolker2020}); \texttt{applefy}\footnote{\url{https://applefy.readthedocs.io/en/latest/}} (\citealt{Bonse2023}).}

\bibliographystyle{aa}
\bibliography{clean_second_revision} %

\appendix

\section{Initial data}

\begin{figure*}
\includegraphics[width=\textwidth]{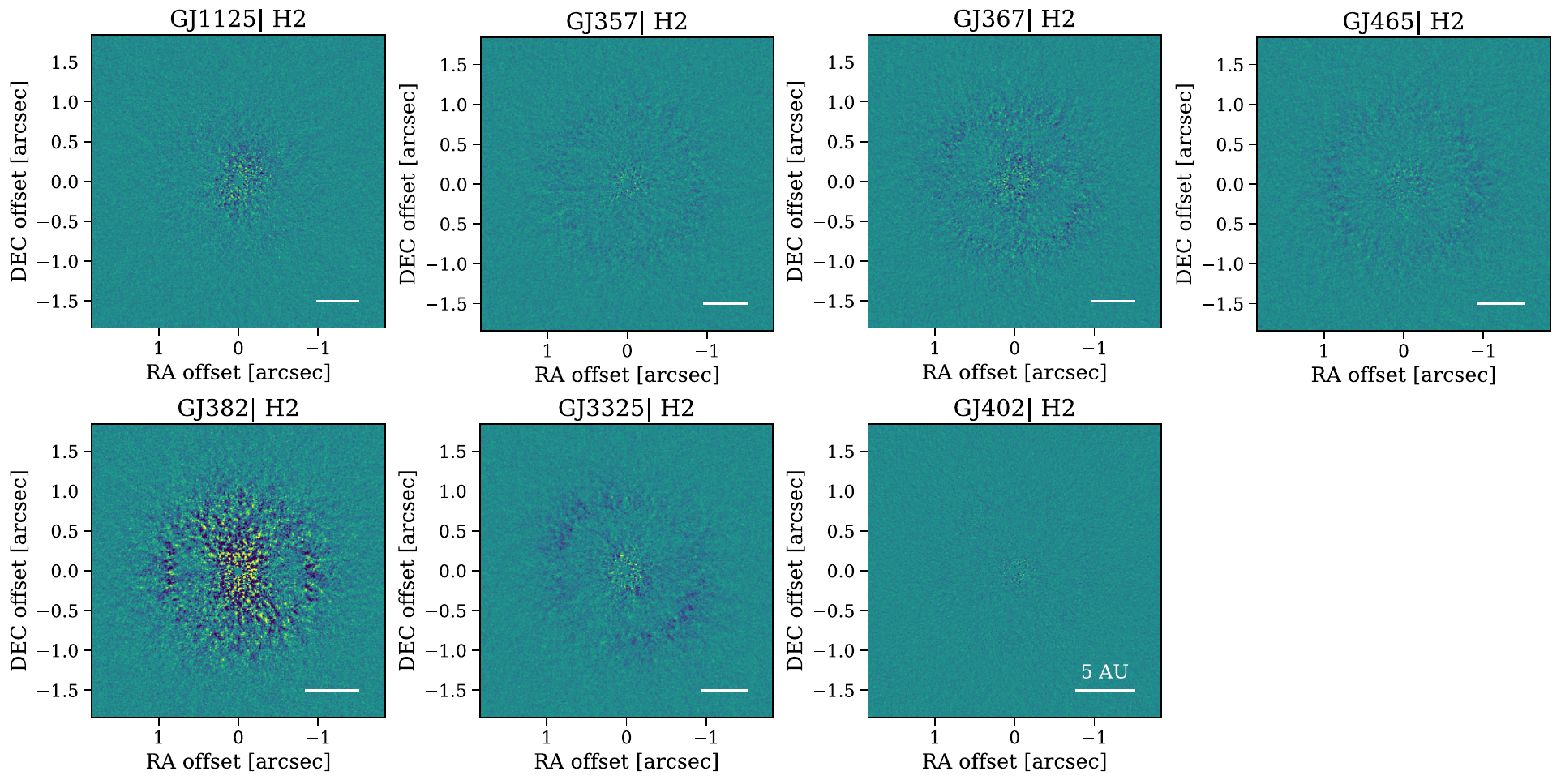}
\caption{Residuals images computed with \texttt{PynPoint} (\citealt{AmaraQuanz2012}; \citealt{Stolker2019}) for the SPHERE/\textit{H2} observations using full-frame PCA and 100 components. The circular feature at $\sim 1$ arcsec corresponds to the limit of the AO correction (0.8 arcsec, \citealt{Fusco2006,Fusco2016}). The white bars correspond to 5 AU at the distance of the star. No companion is detected.}
\label{fig:H2_residuals}
\end{figure*}

\section{Combined statistic plots}\label{app:com}

\begin{figure*}
\includegraphics[width=0.75\textwidth]{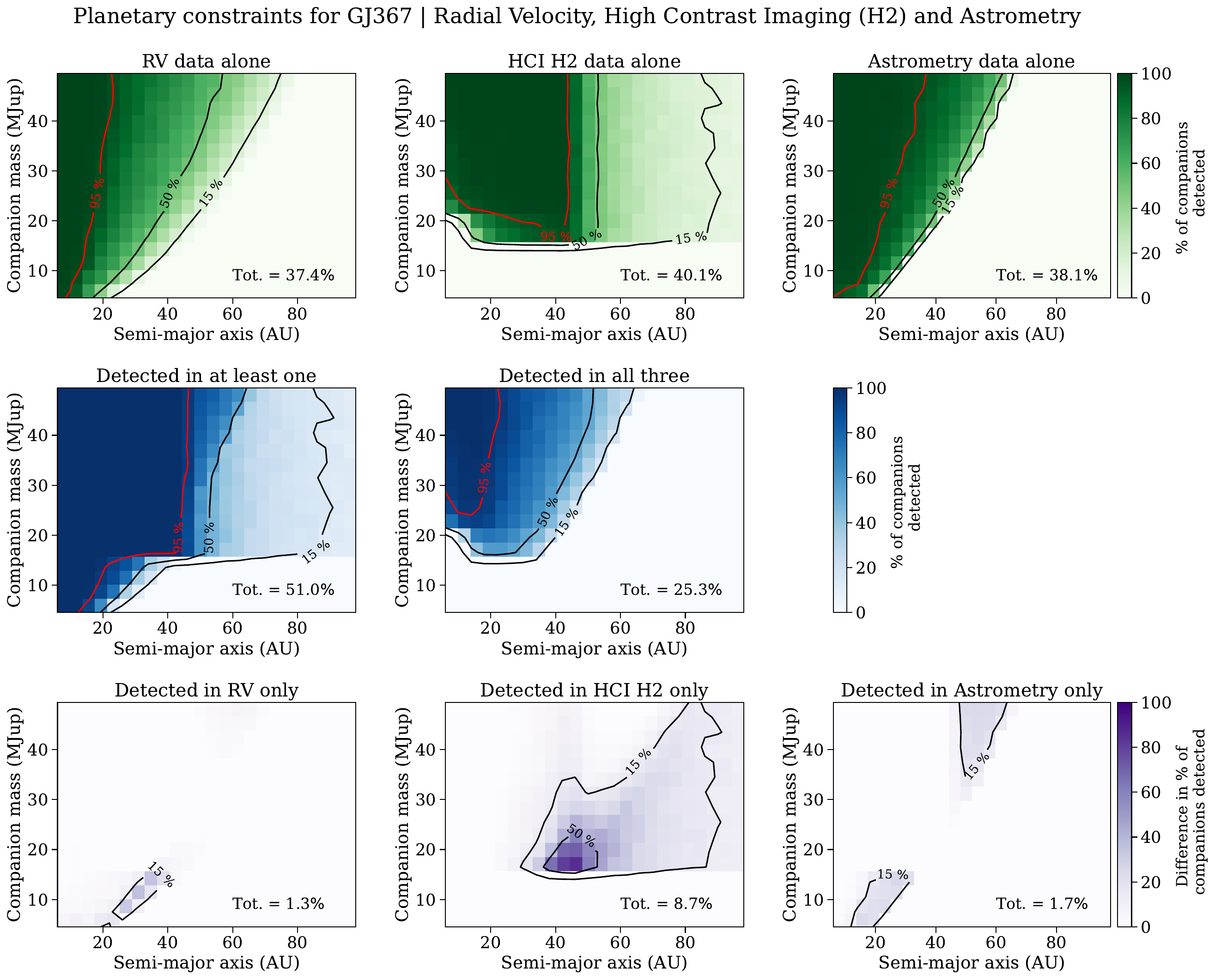}
\caption{Same constraints on companion mass and semi-major axis as shown in \ref{fig:comb_ex_GJ1125_H2} but for GJ367.}
\label{fig:comb_ex_GJ367_H2}
\end{figure*}

\begin{figure*}
\includegraphics[width=0.75\textwidth]{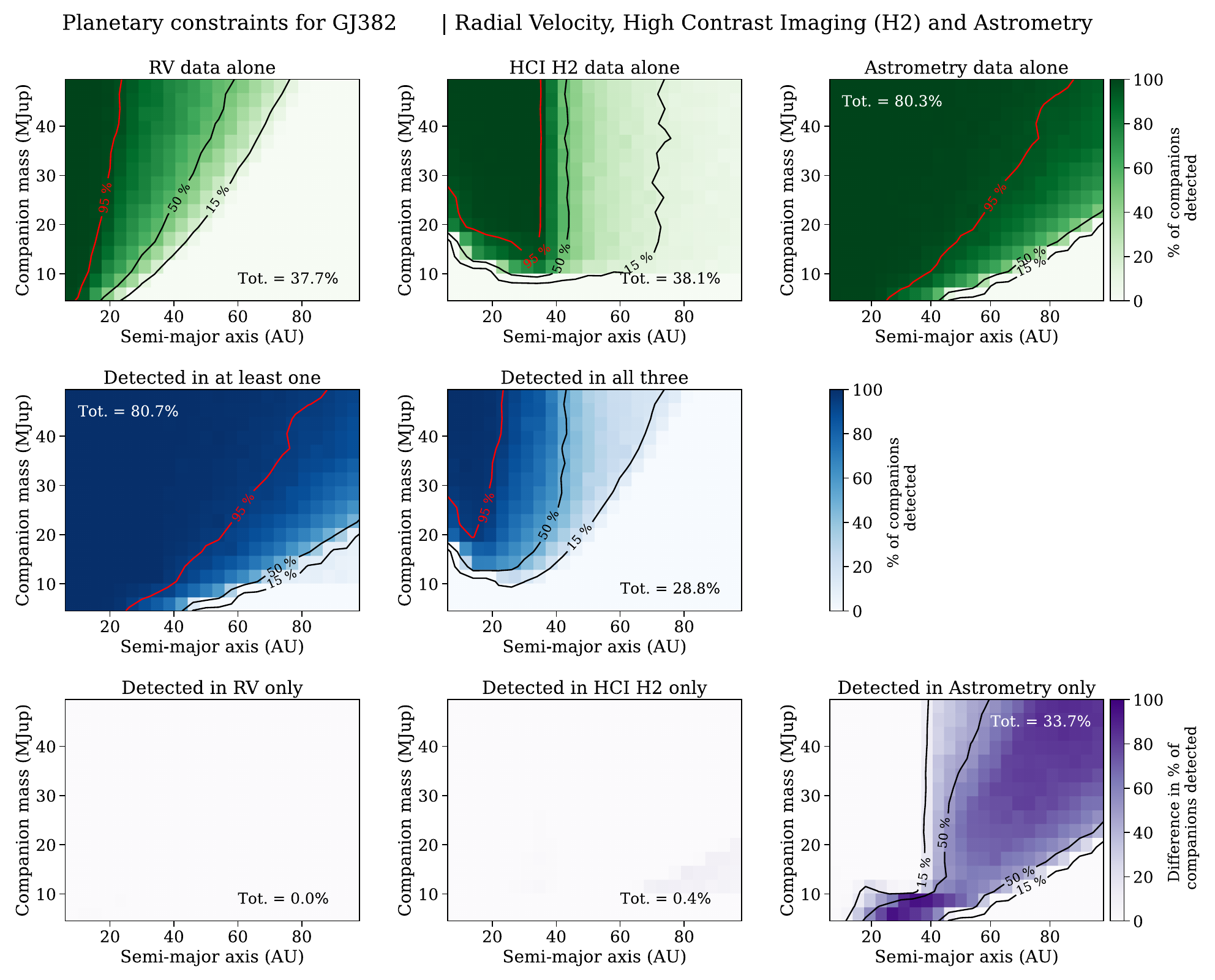}
\caption{Same constraints on companion mass and semi-major axis as shown in Fig. \ref{fig:comb_ex_GJ1125_H2} but for GJ382.}
\label{fig:comb_ex_GJ382_H2}
\end{figure*}

\begin{figure*}
\includegraphics[width=0.75\textwidth]{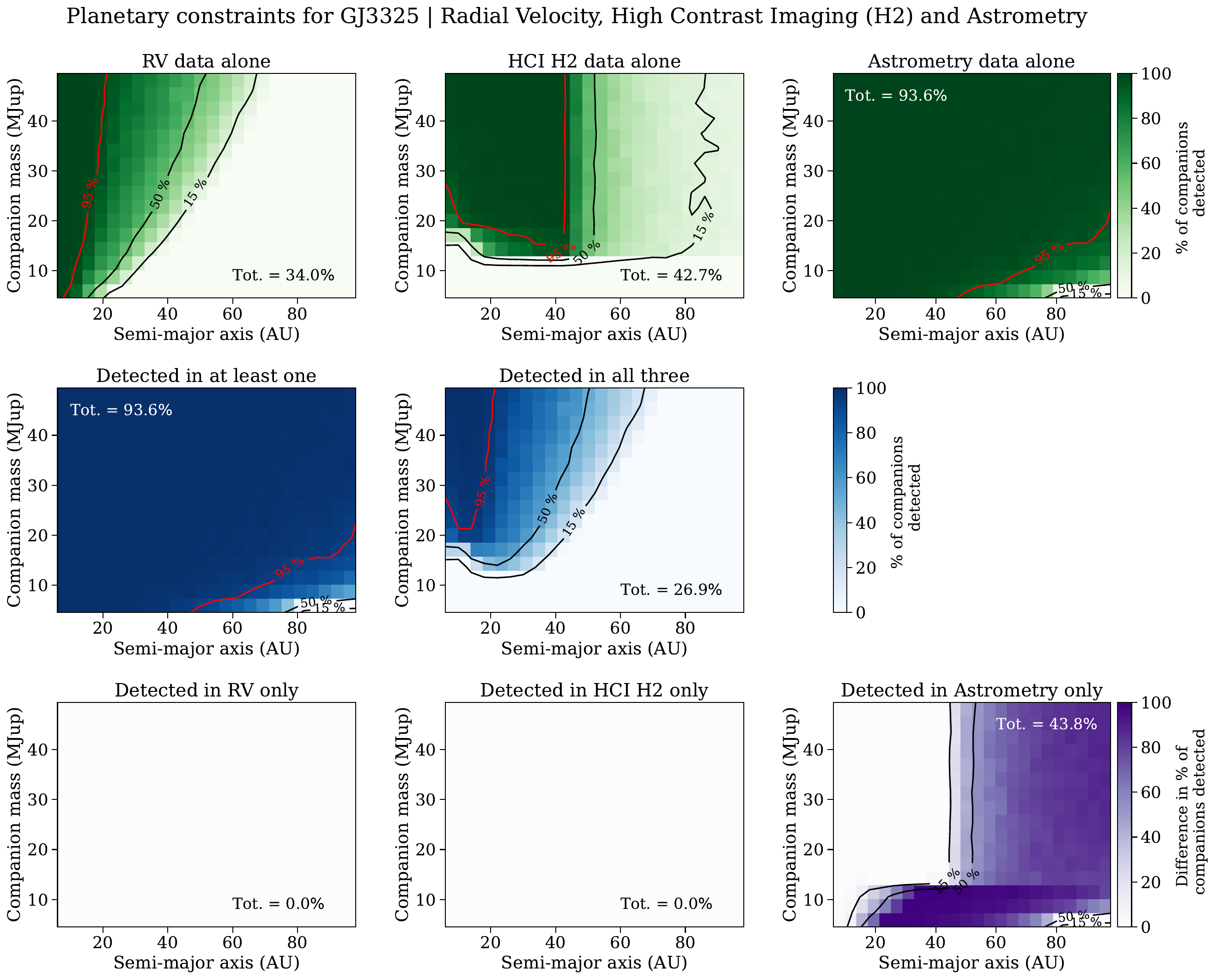}
\caption{Same constraints on companion mass and semi-major axis as shown in Fig. \ref{fig:comb_ex_GJ1125_H2} but for GJ3325.}
\label{fig:comb_ex_GJ3325_H2}
\end{figure*}

\begin{figure*}
\includegraphics[width=0.75\textwidth]{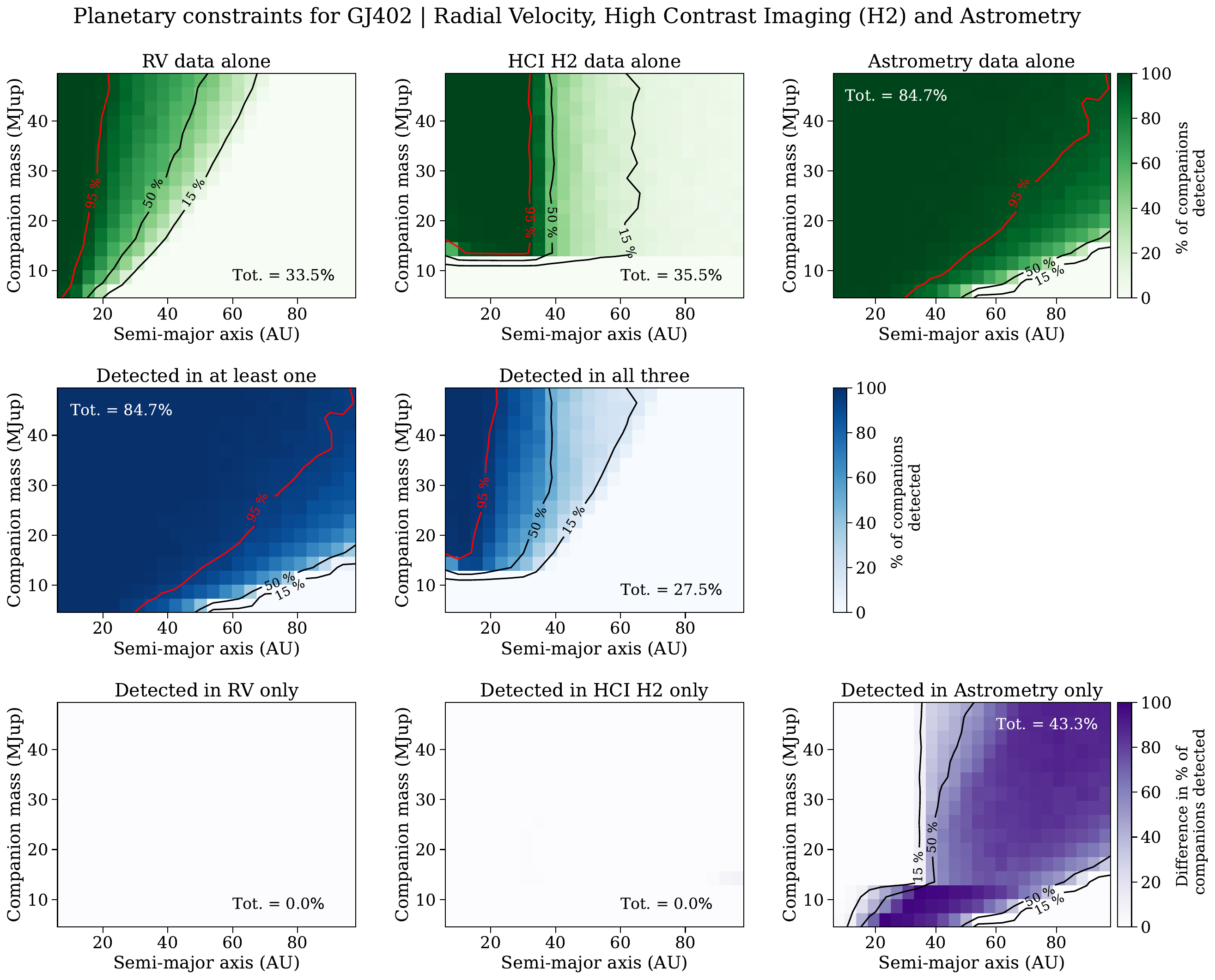}
\caption{Same constraints on companion mass and semi-major axis as shown in Fig. \ref{fig:comb_ex_GJ1125_H2} but for GJ402.}
\label{fig:comb_ex_GJ402_H2}
\end{figure*}

\begin{figure*}
\includegraphics[width=0.75\textwidth]{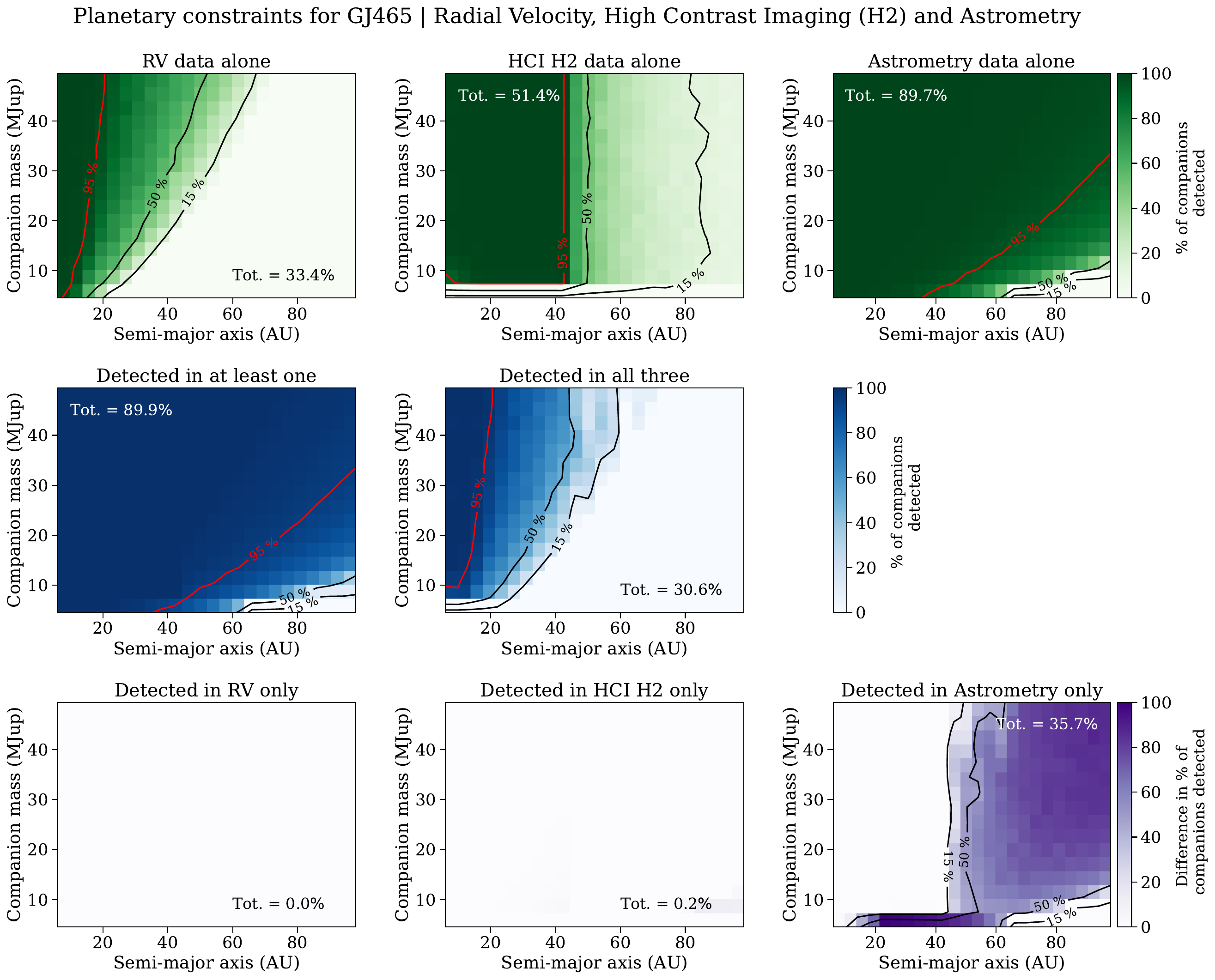}
\caption{Same constraints on companion mass and semi-major axis as shown in Fig. \ref{fig:comb_ex_GJ1125_H2} but for GJ465.}
\label{fig:comb_ex_GJ465_H2}
\end{figure*}

\begin{figure*}
\includegraphics[width=0.75\textwidth]{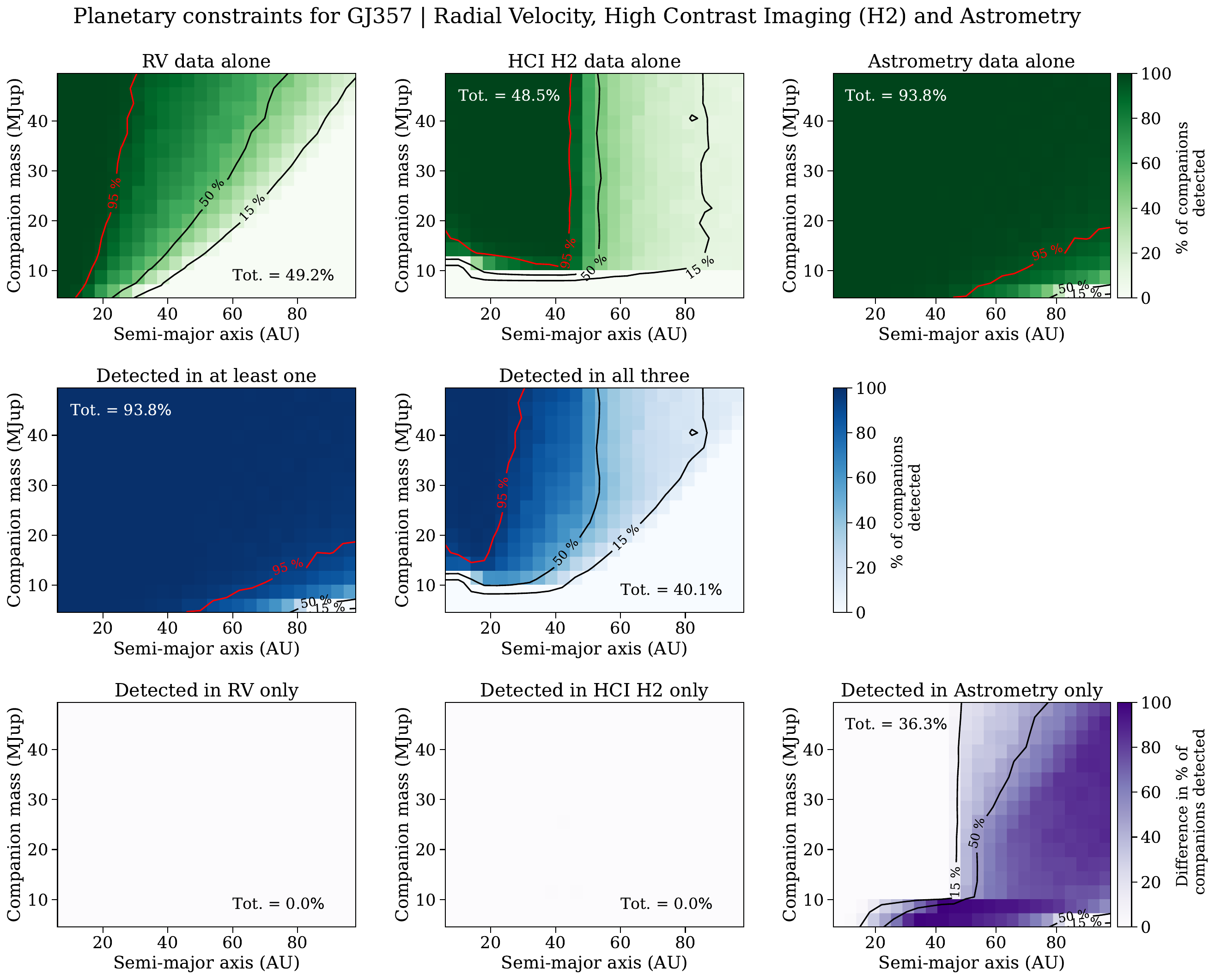}
\caption{Same constraints on companion mass and semi-major axis as shown in Fig. \ref{fig:comb_ex_GJ1125_H2} but for GJ357.}
\label{fig:comb_ex_GJ357_H2}
\end{figure*}

\end{document}